\documentclass[english,a4paper,10pt]{article}
\usepackage{amssymb,amsmath,graphicx}
\usepackage[T1]{fontenc}
\usepackage{tipa}
\usepackage{subfig}
\usepackage{graphicx}
\usepackage{siunitx}
\usepackage{booktabs}
\usepackage[scanall]{psfrag}
\usepackage{comment}
\usepackage{longtable}

\newcommand{\bbm}[1]{\left[\begin{matrix} #1 \end{matrix}\right]}

\newcommand{\GlideFrequencyBand}{[150 \, \mathrm{Hz}, 320 \, \mathrm{Hz} ]}

\newcommand{\RevisionText}[1]{{#1}}
\newcommand{\SecondRevision}[1]{{#1}}
\newcommand{\ThirdRevision}[1]{{#1}}

\usepackage{moreverb,url}

\usepackage[colorlinks,bookmarksopen,bookmarksnumbered,citecolor=red,urlcolor=red]{hyperref}

\newcommand\BibTeX{{\rmfamily B\kern-.05em \textsc{i\kern-.025em b}\kern-.08em
T\kern-.1667em\lower.7ex\hbox{E}\kern-.125emX}}

\title{Modal locking between vocal fold and vocal tract oscillations: Simulations in time domain}

\author{A.~Aalto$^{1,4}$, T.~Murtola$^{1,3}$, J.~Malinen$^{1,*}$, D.~Aalto$^{2,5}$, M.~Vainio$^2$}

\begin{document}
\maketitle

\let\thefootnote\relax\footnote{
\noindent  $^1$Dept. of Mathematics and Systems  Analysis, Aalto University, Finland \\
  $^2$Institute of Behavioural Sciences (SigMe group), 
  University of Helsinki,  Finland\\
  $^3$ Dept. of Signal Processing and Acoustics, Aalto  University, Finland  \\
  $^4$  Luxembourg Centre for Systems Biomedicine, University of Luxembourg, Luxembourg \\ 
  $^5$ Speech Communication and Disorders,  University of Alberta, Canada \\
  $^*$ Corresponding author: jarmo.malinen@aalto.fi}






\begin{abstract}
During voiced speech, the human vocal folds interact with the vocal 
tract acoustics.  The resulting
\RevisionText{glottal source-resonator coupling} has been observed
using mathematical and physical models as well as in {\em in vivo}
phonation.  We propose a computational time-domain model of the full
speech apparatus that, in particular, contains a feedback mechanism
from the vocal tract acoustics to the vocal fold oscillations. It is
based on numerical solution of ordinary and partial differential
equations defined on vocal tract geometries that have been obtained by
Magnetic Resonance Imaging. The model is used to simulate rising and
falling pitch glides of [\textipa{\textscripta , i}] in the
fundamental frequency ($f_o$) interval $\GlideFrequencyBand$.  The
interval contains the first \ThirdRevision{vocal tract resonance $f_{R1}$
and the first formant $F_1$}
of [\textipa{i}] as well as the \ThirdRevision{fractions of the first resonance} $f_{R1}/4$ and $f_{R1}/3$ of [\textipa{\textscripta}].

The simulations reveal a locking pattern of the $f_o$-trajectory at
$f_{R1}$ of [\textipa{i}] in falling and rising glides.  The \ThirdRevision{resonance fractions}
of [\textipa{\textscripta}] produce perturbations in the 
\SecondRevision{pressure signal at the lips} but no locking. All these 
observations from the model
behaviour are consistent and robust within a wide range of feasible
model parameter values and under exclusion of secondary model
components. 

\end{abstract}













\maketitle


\newpage 

\section{\label{IntroSec} Introduction}

The classical source--filter theory of vowel production is built on
the assumption that the source (i.e., the vocal fold vibration)
operates independently of the filter (i.e., the vocal tract,
henceforth VT) whose resonances modulate the resulting vowel sound 
\cite{Chiba:Vowel:1941,Fant:AcoustTh:1960}. Even though this approach 
captures a wide range of phenomena in speech production, at least in male 
speakers, some observations remain unexplained by the source--filter model
lacking feedback. The purpose of this article is to deal with some
of these observations using computational modelling.

More precisely, simulated rising and falling frequency glides of
vowels [\textipa{\textscripta}] and [\textipa{i}] over the frequency
range $\GlideFrequencyBand$ are considered. Similar glides recorded
from eleven female test subjects are treated in the companion
article \cite{A-M-V:MLBVFVTOPII}. Such a vowel glide is particularly
interesting if its glottal frequency ($f_o$) range intersects an
isolated acoustic resonance of the supra- or subglottal cavity, which
we here assume to correspond to the lowest formant $F_1$. Since
$F_1$ almost always lies high above $f_o$ in adult male phonation,
this situation occurs typically in female subjects and only when they
are producing vowels such as [\textipa{i}] with low $F_1$.  As
reported below,
simulations reveal (in addition to other observations) a
characteristic locking behaviour of $f_o$ at the VT acoustic
resonance\footnote{\RevisionText{The VT resonances $f_{R1}, f_{R2},
    \ldots$ are understood here as purely mathematical objects,
    determined by an acoustic PDE and its boundary conditions that are
    defined on the VT geometry.  Formants $F_1, F_2, \ldots$ refer to
    respective frequency peaks extracted from natural \ThirdRevision{or 
    simulated} speech. \ThirdRevision{Here, the notation of 
    \cite{Titze:Notation:2015} is used to differentiate
    the two although,} of
    course, we expect to have $f_{Rj} \approx F_j$ for $j = 1, 2,
    \ldots$.} } 
    $f_{R1} \approx F_1$.  \RevisionText{To check the
  robustness of the model observations, secondary features of the
  model and the role of unmodelled physics are discussed at the end of
  the article.}

As a matter of fact, this article has two equally important
objectives.  Firstly, we pursue better understanding of the
time-domain dynamics of glottal pulse perturbations near $f_{R1}$ of
[\textipa{i}] and at other acoustic ``hot spots'' of the VT and the
subglottal system within $ \GlideFrequencyBand$ that may be reached in
speech or singing. An acoustic and flow-mechanical model of the speech 
apparatus is a \SecondRevision{well-suited} tool for this purpose. Secondly, 
we introduce and validate a computational model that meets these
requirements. The proposed model has been originally designed to be a
glottal pulse source for high-resolution 3D computational acoustics
model of the VT which is being developed for medical purposes
\cite{Aalto:MesTec:2014}.
There is an emerging application for this model as a development
platform of speech signal processing algorithms such as discussed in
\cite{Alku:IAIF:1992}, \cite{Alku:IFReview:2011} and
\cite{Alku:FEst:2013}; however, the model introduced in
\cite{Story:1995} has been used in \cite{Alku:FEst:2013}. Since
perturbations of $f_o$ near $F_1$ are a widely researched, yet quite
multifaceted phenomenon, it is a good candidate for model validation
experiments.

\SecondRevision{The simulations carried out in this paper} indicate 
special kinds of perturbations in vocal folds
vibrations near a VT resonance \RevisionText{as reported below.}
The mere existence of such perturbations is \SecondRevision{not} surprising 
considering the wide range of existing literature. Since the seminal 
work of \cite{Ishizaka:1972}, 
\RevisionText{a wide range of glottal source perturbation patterns
  related to acoustic loading has been investigated. Experiments were
  carried out in \cite{Austin:SGR:1997} on excised larynges mounted on
  a resonator to determine how glottal amplitude ratio changes with
  the subglottal resonator length. Physical models were used in
  \cite{Zhang:SGAcoust:2005} with a subglottal resonator to study
  phonation onsets and offsets, and in \cite{Lucero:2012} with sub-
  and supraglottal resonators to study phonation onsets. The latter
  also considered the dynamics of frequency jumps as the natural
  frequency of their physical model was varied over time. Similarly, a
  physical model of phonation with tubular, variable length
  supraglottal resonator was studied in \cite{Ruty:2007,Ruty:2008},
  and it was used to validate a flow-acoustic model somewhat
  resembling the one proposed in this article.
  
  In \cite{Titze:NonlinCouplingTh:2008} the problem was approached 
  using both reasoning based on sub- and supraglottal impedances and a
  non-computational flow model as well as computational model
  comprising a multi-mass vocal fold model and wave-reflection models
  of the subglottal and supraglottal systems.}
A two-mass model of vocal folds, coupled with a variable-length resonator 
tube, was used in \cite{Hatzikirou:VoiceInstabil:2006}, and pitch glides 
were simulated using a four-mass model to analyse the interactions between 
vocal register transitions and VT resonances in
\cite{Tokuda:RegisterTransitions:2010}. 

These works reveal a consistent picture of the existence of perturbations 
\RevisionText{caused by resonant loads}, and this phenomenon has also been detected
experimentally in \cite{Titze:NonlinCouplingExp:2008} using speech
recordings and in \cite{Zanartu:2011} using simultaneous recordings of
laryngeal endoscopy, acoustics, aerodynamics, electroglottography, and
acceleration sensors. The latter article also contains a review on
related voice bifurcations.

\SecondRevision{Although the existence of these perturbations has been
  well reported, speech modelling studies have given only limited
  attention to the time-domain dynamics of fundamental frequency
  glides where such perturbations would be expected to occur. Of the
  above mentioned studies, upward glides were simulated in
  \cite{Lucero:2012} by varying the natural frequency of their
  physical model over time.  Their small amplitude oscillation model
  exhibited a frequency jump in the vicinity of the resonance of their
  downstream tube when the load resistance was sufficiently strong.
  Downward glides were simulated in \cite{Titze:NonlinCouplingTh:2008}
  followed by upward glides by varying the parameters of a multi-mass
  vocal fold model. Frequency jumps, subharmonics and amplitude
  changes were observed in the regions where load reactances were
  changing rapidly. Changes in the rate of change of the fundamental
  frequency in these regions can also be seen in their Figures
  10-14. In \cite{Tokuda:RegisterTransitions:2010} upward glides were
  simulated followed by downward glides by adjusting the tension
  parameter (i.e. decreasing masses and increasing stiffness
  parameters by the same factor) in their four-mass vocal fold
  model. They observed frequency jumps associated with register
  changes, which in turn were shown to occur at different frequencies
  depending on the vocal tract load.}

\SecondRevision{Some of the most popular approaches to modelling 
phonation are} based on the Kelly--Lochbaum
vocal tract \cite{Kelly:1962} or various transmission line analogues
\cite{Dunn:CVR:1950,ElMasri:DTL:1998,Mullen:WPM:2006}. Contrary to
these approaches, the proposed model consists of (ordinary and
partial) differential equations, conservation laws, and coupling
equations. In this modelling paradigm, the temporal and spatial
discretisation is conceptually and practically separated from the
actual mathematical model of speech. The computational model is simply
a numerical solver for the model equations, written in MATLAB
environment. The modular design makes it easy to decouple model
components for assessing their significance to simulated
behaviour.\footnote{Some economy of modelled features should be
  maintained to prevent various forms of ``overfitting'' while
  explaining the experimental facts. Good modelling practices within
  mathematical acoustics have been discussed in Chapter~8 in
  \cite{Rienstra:Acoust:2013}.} Since the generalised Webster's
equation for the VT acoustics assumes intersectional area functions as
its geometric data, VT configurations from Magnetic Resonance Imaging
(MRI) can be used without transcription to non-geometric model
parameters. Thus, time-dependent VT geometries are easy to
implement. Further advantages of speech modelling based on Webster's
equation have been explained in \cite{Doel:Webster:2008} where the
approach is somewhat similar to one taken here.

The proposed model aims at qualitatively realistic functionality,
tunability by a low number of parameters, and tractability of model
components, equations, and their relation to biophysics.  Similar
functionality in higher precision can be obtained using Computational
Fluid Dynamics (CFD) with elastic tissue boundaries. In the CFD
approach, the aim is to model the speech apparatus as undivided whole
\cite{Horacek:Airflow:2011},
but the computational cost is much higher compared to our model or the
models proposed in, e.g., \cite{Doel:Webster:2008} and
\cite{Ho:SGT:2011}. The numerical efficiency is a key issue because
some parameter values or their feasible ranges (in particular, for
hard-to-get physiological parameters) can only be determined by the
trial and error method \RevisionText{as discussed in Chapter~4 in
  \cite{Murtola:MVP:2014}, leading to a high number of required
  simulations.}

\section{\label{VocalFoldsSec} Model of the Vocal Folds}

\subsection{Anatomy, physiology, and control of phonation}

All voiced speech sounds originate from self-sustained quasi-periodic
oscillations of the vocal folds where the closure of the aperture ---
known as the rima glottidis --- between the two string-like vocal
folds cuts off the air flow from lungs. This process is called
phonation, and the system comprising the vocal folds and the rima
glottidis is known as the glottis. A single period of the glottal flow
produced by phonation is known as a glottal pulse. \RevisionText{A
  description of structures in human larynx and their function can be
  found, e.g., in \cite{stevens2000} or \cite{hirose2010investigating},
  and we give here only a brief summary.}

\begin{figure}[!tbp]
  \centering
  \begin{minipage}[b]{0.4\textwidth}
      \includegraphics[width=4cm]{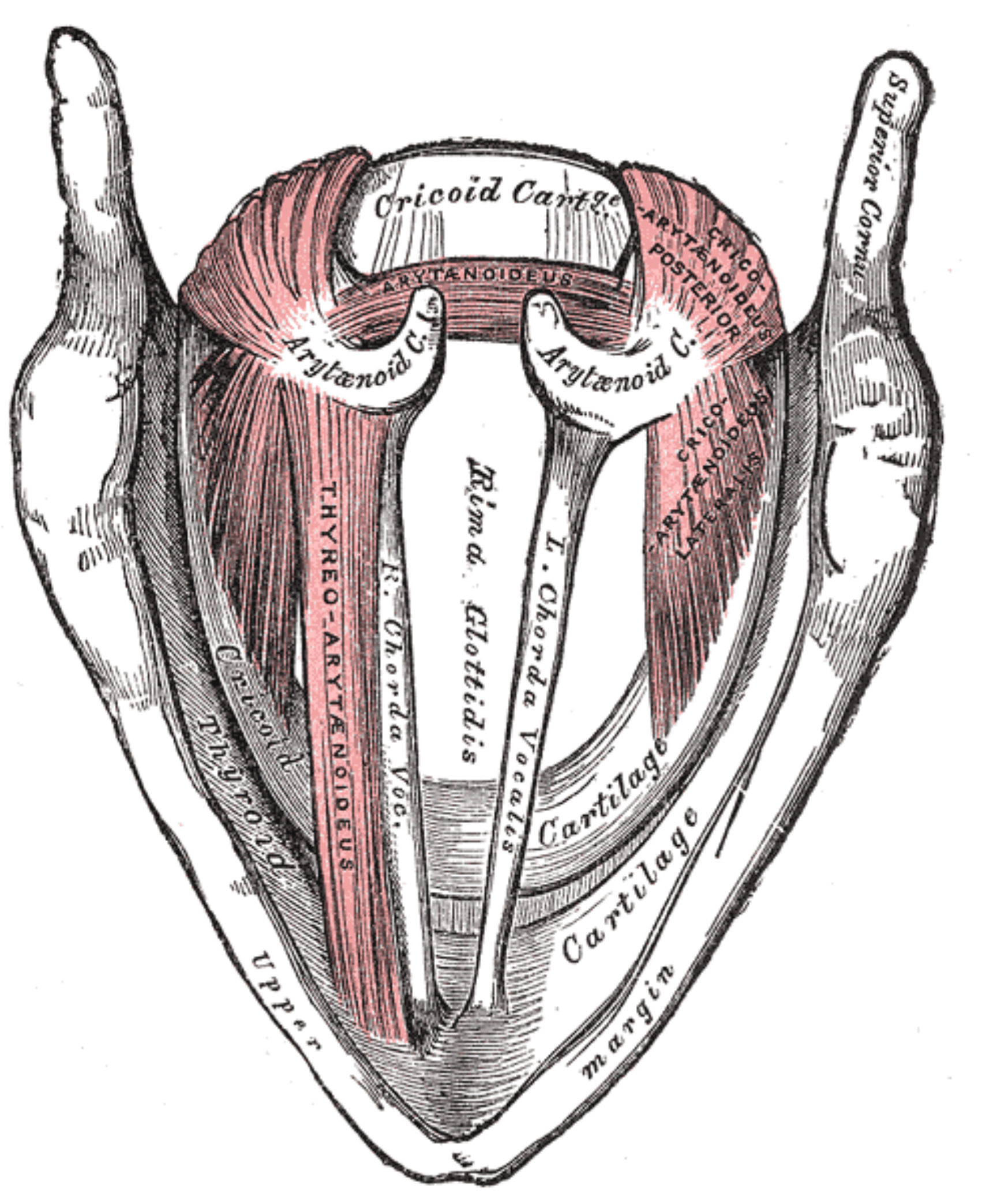} 
      \includegraphics[width=5.3cm]{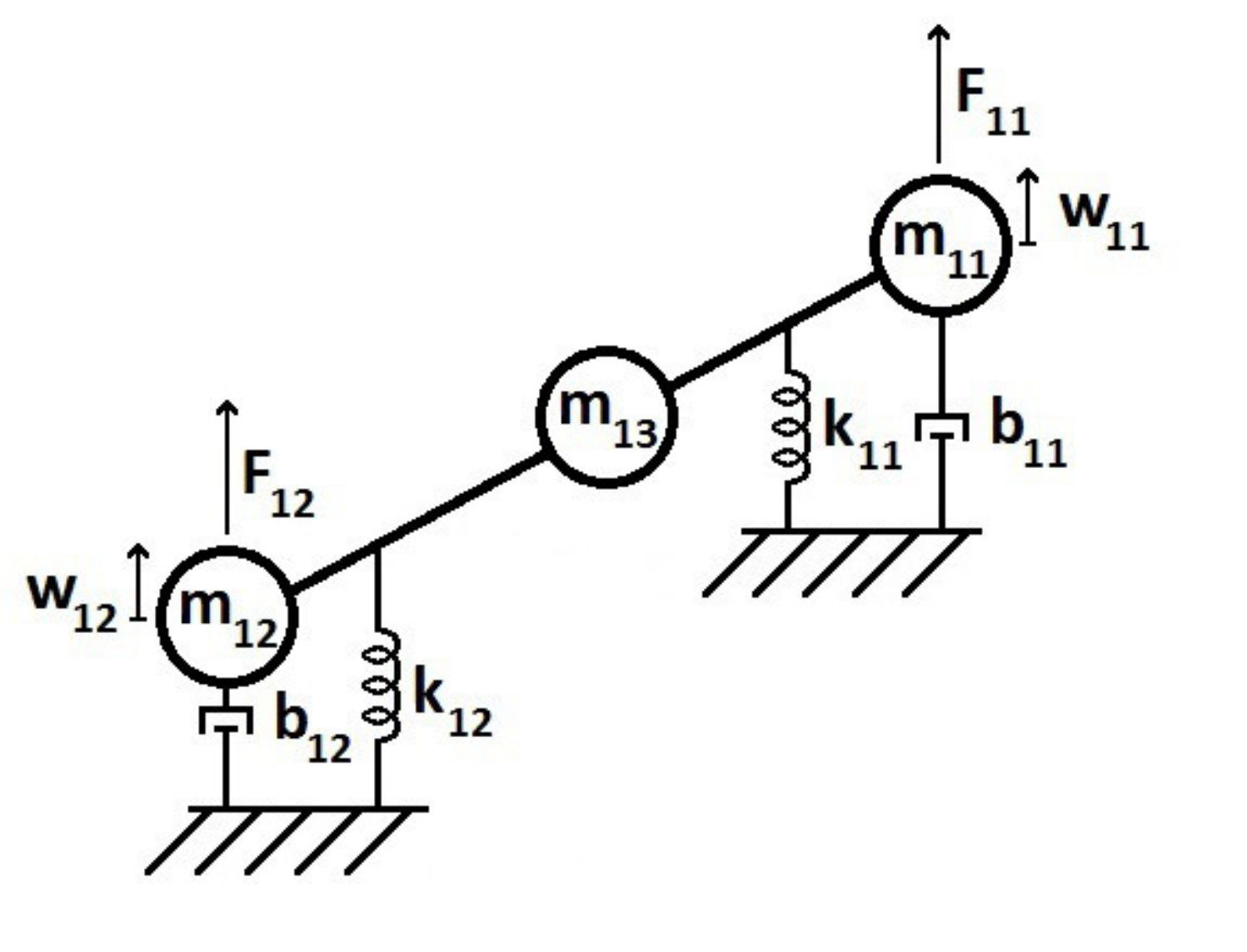}
  \end{minipage}
  \begin{minipage}[b]{0.4\textwidth}
      \includegraphics[width=7cm]{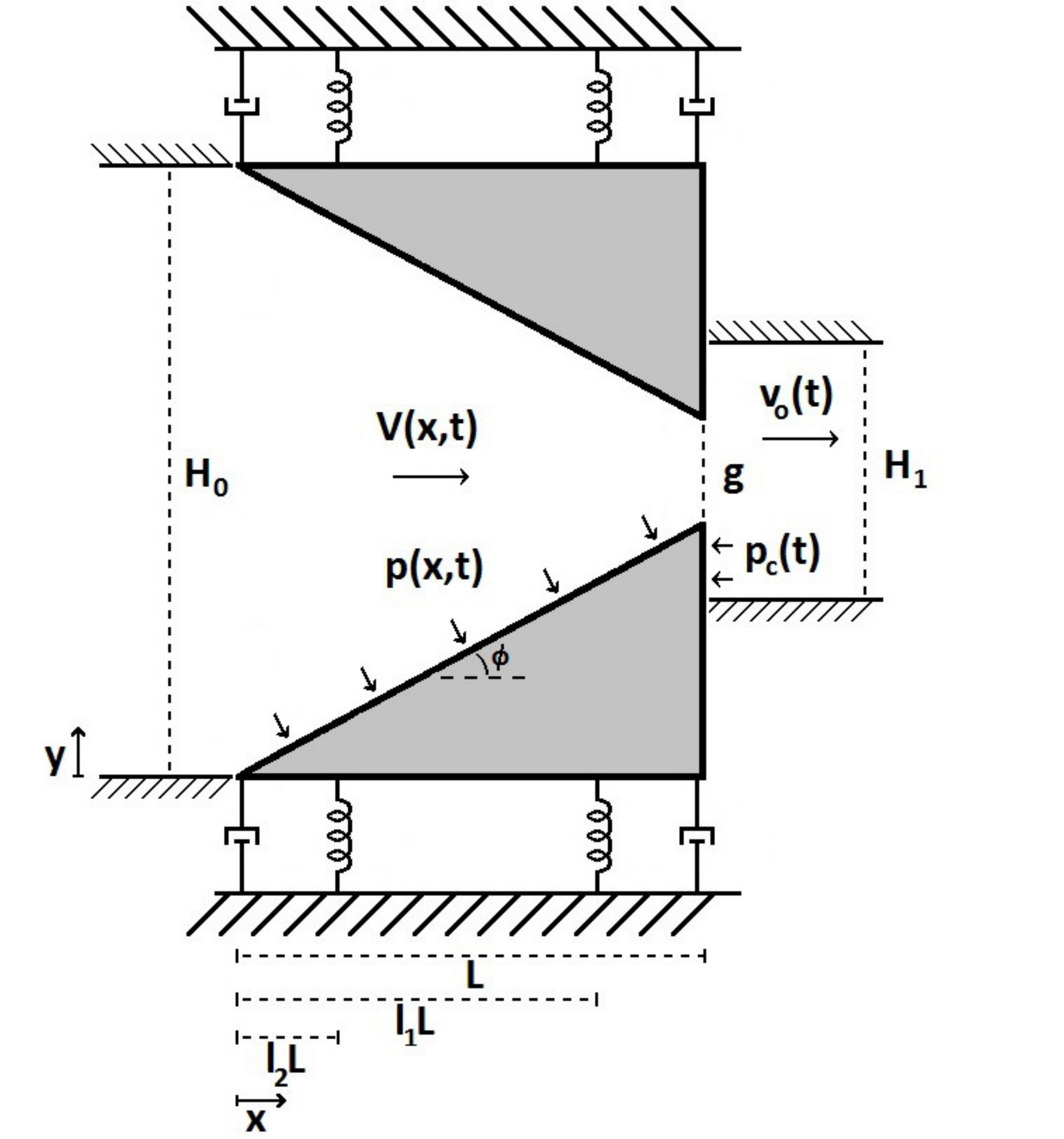}
  \end{minipage}
\caption{The topmost panel on left: Sketch of the anatomy of the
  larynx seen from above according to \protect \cite{Gray:1918}. Right
  panel: The geometry of the glottis model and the symbols used. The
  trachea (i.e., the channel leading from the lungs to glottis) is to
  the left in this sketch and the vocal tract is to the right.  The
  lower panel on left: \SecondRevision{Lumped-element representation of the glottis 
  model with two degrees of freedom shown for the lower vocal fold ($j=1$).}}
   \label{fig:model}
\end{figure}

 


As shown in Figure~\ref{fig:model} (\SecondRevision{upper} left panel), 
each vocal fold consists of a vocal ligament (also known as a vocal cord) 
together with a medial part of the thyroarytenoid muscle, and the vocalis 
muscle \SecondRevision{(not specified in upper left panel of} Figure~\ref{fig:model}). Left and right vocal folds are attached 
to the thyroid cartilage from their anterior ends and to the respective 
left and right arytenoid cartilages from their posterior ends. In addition, 
there is the fourth, ring-formed cricoid cartilage whose location is 
inferior to the thyroid cartilage. The vocal folds and the associated 
muscles are supported by these cartilages.

There are two muscles attached between each of the arytenoid
cartilages and the cricoid cartilage: the posterior and the lateral
cricoarytenoid muscles whose mechanical actions are opposite. The vocal 
folds are adducted by the contraction of the lateral cricoarytenoid 
muscles during phonation, and conversely, abducted by the
posterior cricoarytenoid muscles during, e.g., breathing. This control
action is realised by a rotational movement of the arytenoid
cartilages in a transversal plane. In addition, there is a fifth
(unpaired) muscle --- the arytenoid muscle --- whose contraction
brings the arytenoid cartilages closer to each other, thus reducing
the opening of the glottis independently of the lateral cricoarytenoid
muscles. These rather complicated control mechanisms regulate the type
of phonation in the breathy-pressed scale.

The main mechanism controlling the fundamental frequency $f_o$ of
voiced speech sound is actuated by two cricothyroid muscles (not
visible in \SecondRevision{upper left panel of} Figure~\ref{fig:model}). The contraction of these muscles
leads to a rotation of the thyroid cartilage with respect to the
cricoid cartilage.  As a result, the thyroid cartilage inclines to the
anterior direction, thus stretching the vocal folds. The elongation of
the string-like vocal folds leads to increased stress which raises the
fundamental frequency $f_o$ of their longitudinal vibrations. The
vertical movement of larynx also rotates cricoid cartilage impacting
$f_o$. Finally, the phonation and $f_o$ are influenced by subglottal
pressure through the control of respiratory muscles.

\subsection{Glottis model}

The anatomic configuration in \ThirdRevision{upper left panel of}
Figure~\ref{fig:model} is idealised as a low-order mass-spring system
with aerodynamic surfaces as shown in \ThirdRevision{right panel of}
Figure~\ref{fig:model} and discussed in \cite{Aalto:MT:2009} and
\cite{Murtola:MVP:2014}.
Such lumped-element models have been used frequently (see, e.g.,
\cite{Horacek:2005}, \cite{Liljencrants:1991}, \cite{Lous:1998},
\cite{Pelorson:FlowSepar:1994}, \cite{Ruty:2008}, and \cite{Story:Body-cover:1995}) 
since the introduction of the classic two-mass model \cite{Ishizaka:1972}.
For recent reviews of the variety of lumped-element
and PDE based models and their applications, see
\cite{Alipour:MMNS:2011}, \cite{Birkholz:LumpedElementReview:2011} and
\cite{Erath:LumpedElementReview:2013}.

The radically simplified glottis model geometry in
Figure~\ref{fig:model} (right panel) corresponds to the coronal
section through the center of the vocal folds. Both the fundamental
frequency $f_o$ as well as the phonation type can be chosen by
adjusting parameter values (see Section~4 in
\cite{Murtola:MVP:2014}). Register shifts (e.g., from modal register to
falsetto) are not in the scope of this model since it would require
either modelling the vocal folds as aerodynamically loaded strings or
as a high-order mass-spring system that has a string-like ``elastic''
behaviour.

The vocal fold model in Figure~\ref{fig:model} (right panel) consists
of two wedge-shaped moving elements that have two degrees of freedom
each: \SecondRevision{each end of the vocal fold can move in the 	
  y-direction. Although this causes some distortion to the shape of 
  the wedges, the displacements are small enough that this effect is 
  negligible.} 
The distributed mass of these elements \ThirdRevision{is} reduced into
three mass points which, \SecondRevision{for the $j^{\textrm{th}}$
  fold, $j = 1,2$,} are located so that $m_{j1}$ is at $x=L$, $m_{j2}$
at $x=0$, and $m_{j3}$ at $x=L/2$.  Here $L$ denotes the thickness of
the modelled vocal fold structures. \ThirdRevision{In calculation of
  the masses, the reduced system retains the mass, and static and
  inertial moments of a more realistic vocal fold shape}
\SecondRevision{(for details, see \cite{Aalto:MT:2009} p. 14)}. The
elastic support of the vocal ligament is approximated by two springs
at \SecondRevision{points $x=l_1L$ and $x=l_2L$.} The equations of
motion for the vocal folds are given by
\begin{equation} 
\begin{cases}
  \  M_1\ddot{W}_1(t)+B_1\dot{W}_1(t)+K_1W_1(t)=F_1(t), \\
  \ M_2\ddot{W}_2(t)+B_2\dot{W}_2(t)+K_2W_2(t)=F_2(t), \qquad t \in
  \mathbb{R},
\end{cases} \label{eq:liikeyhtalot}
\end{equation}
where $W_j= \bbm{w_{j1} & w_{j2}}^T$ are the displacements of 
	\SecondRevision{$m_{j1}$ and $m_{j2}$ in the y-direction} as shown in 
	Figure~\ref{fig:model} 
	(\SecondRevision{lower left} panel). \RevisionText{The
  loading force pair $F_j(t) = \bbm{F_{j1}(t) & F_{j2}(t)}^T$ is due to
  aerodynamic \SecondRevision{and acoustic pressure} forces in Eq.~\eqref{eq:aeroforces} when the glottis is
  open, and the collision forces in Eq.~\eqref{HertzImpactEq} when the
  vocal folds are in contact.}  The respective mass, damping, and
stiffness matrices $M_j$, $B_j$, and $K_j$ in \eqref{eq:liikeyhtalot}
are
\begin{equation} \label{eq:matrices}
\begin{array}{c}
M_j  =  \left[ \begin{array}{cc} m_{j1}+\frac{m_{j3}}{4} & \frac{m_{j3}}{4} \\
\frac{m_{j3}}{4} & m_{j2}+\frac{m_{j3}}{4} \end{array} \right],
\hspace{6mm} 
B_j = \left[ \begin{array}{cc} b_{j1} & 0 \\ 0 & b_{j2} \end{array} \right], \vspace{4mm}\\
\textrm{and } \quad K_j = \left[ \begin{array}{cc} l_1^2k_{j1}+l_2^2k_{j2} & l_1l_2(k_{j1}+k_{j2}) \\ l_1l_2(k_{j1}+k_{j2}) & l_2^2k_{j1}+l_1^2k_{j2} \end{array} \right]. \end{array}
\end{equation}
The entries of these matrices have been computed by means of
Lagrangian mechanics \RevisionText{in \cite{Aalto:MT:2009}}.  The
damping matrices $B_j$ are diagonal since the dampers are located at
the endpoints of the vocal folds. The model supports asymmetric vocal
fold vibrations but for this work \SecondRevision{symmetry is imposed 
	on the vocal folds by using parameters $M=M_j$, $K=K_j$, and $B=B_j$, 
	$j =1,2$, and by setting $F(t)= F_2(t)= -F_1(t)$}. \RevisionText{The parameters in equation~\eqref{eq:matrices} as
  well as the loading force components in
  equation~\eqref{eq:liikeyhtalot} are illustrated in
  Figure~\ref{fig:model}} \SecondRevision{(right and lower left panels)}.

The glottal openings at the two ends of the vocal folds, denoted by
$\Delta W_i$, $i = 1,2$, are related to equations~\eqref{eq:liikeyhtalot} through
\begin{equation}
\left[ 
 \begin{array}{c} \Delta W_1 \\
    \Delta W_2 \end{array}
\right]
=W_2-W_1+\left[
\begin{array}{c} g \\ H_0 \end{array} \right]
\end{equation}
where the rest gap parameters $g$ and $H_0$ are as in
Figure~\ref{fig:model} (right panel). In human anatomy, the parameter
$g$ is related to the position and orientation of the arytenoid
cartilages. 

As is typical in related biomechanical modelling
\cite{Horacek:2005,Horacek:PhonThreshold:2002,Tokuda:RegisterTransitions:2010},
the lumped parameters of the mass-spring system
\eqref{eq:liikeyhtalot}-- \eqref{eq:matrices} are in some
correspondence to the true masses, material parameters, and geometric
characteristics of the sound producing tissues.  More precisely,
matrices $M$ correspond to the vibrating masses of the vocal folds,
including the vocal ligaments together with their covering mucous
layers and (at least, partly) the supporting vocalis muscles.  The
elements of the matrices $K$ are best understood as linear
approximations of $k(s) = f/s$ where $f = f(s)$ is the contact force
required for deflection $s$ at the center of the string-like vocal
ligament in the anatomy shown in Figure~\ref{fig:model}
(\SecondRevision{upper} left panel).  It should be emphasised that the
exact numerical correspondence of tissue parameters to lumped model
parameters $M$ and $K$ is intractable (and for most practical purposes
even irrelevant), and their values in computer simulations must be
tuned using measurement data of $f_o$ and the measured form of the
glottal pulse \cite{Murtola:MVP:2014}.

\subsection{Forces during the closed phase}
\RevisionText{During the glottal closed phase (i.e., when $\Delta
  W_1(t) <0$ at the narrow end of the vocal folds), there are no
  proper aerodynamic forces affecting the vocal folds dynamics in
  equations \eqref{eq:liikeyhtalot}.  There are, however, nonlinear
  spring forces with parameter $k_H$, accounting for the elastic
  collision of the vocal folds.  They are accompanied by the resultant
  acoustic counter pressure from the VT and subglottal cavities,
  denoted by $p_c = p_c(t)$ in equation~\eqref{eq:pc} below.}  Thus,
the force pair for equations \eqref{eq:liikeyhtalot} during glottal
\SecondRevision{closed phase} is given by
\begin{equation} \label{HertzImpactEq}
  F = F_H =\left[ \begin{array}{c} k_H|\Delta W_1|^{3/2}- C_{pc} p_c\\
      C_{pc} p_c
\end{array} \right] \quad \mathrm{for} \quad \Delta W_1< 0.
\end{equation}
\SecondRevision{Here, the coupling coefficient $C_{pc}=C_{pc}(t)$ 
accounts for the moment arms and areas on which $p_c$ acts, 
and it will be given an expression in equation \eqref{eq:pc_coeff}.}

This approach is related to the Hertz impact model that has been used
similarly in \cite{Horacek:2005} and \cite{Zanartu:AL1MM:2007}.
\RevisionText{During the glottal open phase (i.e., when $\Delta W_1(t)
  > 0$), the spring force in equations \eqref{HertzImpactEq} is not
  enabled.  Then the load terms in equation~\eqref{eq:liikeyhtalot}
  are given by $F(t)=F_A(t)$ as introduced below in
  equation~\eqref{eq:aeroforces} in terms of the aerodynamic forces
  from the glottal flow.}

\section{\label{FlowSec}Glottal Flow and the Aerodynamic Force}

\SecondRevision{The air flow within the glottis is assumed to be 
incompressible and one-dimensional with velocity $V(x,t)$,} 
satisfying the mass conservation law $H(x,t)V(x,t)=H_1v_o(t)$, where 
$H(x,t)$ is the height of the flow channel inside the glottis. In 
the model geometry of Figure~\ref{fig:model} (right panel) we have
\begin{equation}
  H(x,t)=\Delta W_2(t)+\frac{x}{L}(\Delta W_1(t)-\Delta W_2(t)), \ \ \ x \in [0,L]. \nonumber
\end{equation}

\SecondRevision{The velocity $v_o(t)$ of the flow through
the control area $h H_1$ superior to the glottis is} described by
 \begin{equation} \label{eq:vout}
	\dot{v}_o(t) = \frac{1}{C_{iner} h H_1} \left(p_{sub} - 
 	R_g(t) v_o(t) \right)
 \end{equation}
 where $p_{sub}$ is \ThirdRevision{an ideal pressure source (whose
   values are given relative to the ambient air pressure) located
   immediately inferior to the vocal folds}, $C_{iner}$ regulates flow
 inertia, $h$ is the width of the rectangular flow channel, and
 $R_g(t)$ represents the total pressure loss in the glottis.  In fact,
 equation~\eqref{eq:vout} is related to Newton's second law for the
 air column in motion.  \ThirdRevision{Although the pressure driving
   phonation originates at the lungs, it is here assumed that
   physiological mechanisms enable the adequate control of the
   pressure source $p_{sub}$.}  \SecondRevision{Note that due to
   assumed incompressibility, equation~\eqref{eq:vout} can
   equivalently be written in terms of position-independent volume
   flow through the glottis $U(t) = v_o(t)h H_1$}.

\SecondRevision{To derive equation~\eqref{eq:vout} following
  \cite[Section 2.2]{Aalto:MT:2009}, one begins with the pressure
  loss balance $p_{sub} = p_g + p_a$ where the $p_{sub}$ is the sum of
  the glottal pressure loss and the accelerating pressure of the fluid
  column mass in the airways and lungs. The power of accelerating or
  decelerating an (incompressible) fluid column is $p_a(t) (h H_1)
  v_0(t)$. This power is equal to the derivative of the kinetic energy
  of the fluid column, yielding the identity $p_a(t) (h H_1) v_0(t) =
  \rho v_0(t) v_0'(t) (h H_1)^2 \int{ \frac{d \vec r}{A(\vec r)^2} }$
  where the integration is extended over the VT and SGT volume. Here
  $A(\vec r)$ denotes the area of the fluid column cross-section that
  contains the position vector $\vec r$, and the incompressibility
  $A(\vec r) v(\vec r, t) = h H_1 v_0(t)$ was used.
  Equation~\eqref{eq:vout} follows from this by denoting $C_{iner} =
  \rho \int{ \frac{d \vec r}{A(\vec r)^2}
  }$.} \ThirdRevision{The contribution of the VT to the total inertance
  can be further integrated to $C_{iner}^{VT}=\rho \int_0^{L_{VT}}{
   \frac{ds}{A(s)}}$ but the inertance of the subglottal masses
    cannot be expressed similarly in terms of anatomic data. Hence,
    the parameter $C_{iner}$ has to be used as a tuning parameter.}

 The total pressure loss \SecondRevision{in the glottis} in 
 equation~\eqref{eq:vout} consists of two components, namely
 \begin{equation}  \label{eq:rtot}
\begin{aligned}
 	R_g(t) & = R_v(t) + R_t(t), \quad  \text{ where }  \\ 
         R_v(t) & = \frac{12 \mu H_1 L_g}{\Delta W_1(t)^3} \quad \text{ and } \quad R_t(t) = k_g
        \frac{\rho H_1^2 v_o(t)}{2\Delta W_1(t)^2}.
\end{aligned}
 \end{equation}
The first term $R_v(t)$ represents the viscous pressure loss, and it
is motivated by the Hagen--Poiseuille law in a narrow aperture.  It
approximates the pressure loss in the glottis using a rectangular tube
of width $h$, height $\Delta W_1$, and length $L_g$.  The parameter
$\mu$ is the kinematic viscosity of air.  The second term $R_t(t)$
takes into account the pressure losses not attributable 
to viscosity in the sense of $R_v$, and its form
is motivated by the experimental work in \cite{Berg:1957}.
The coefficient $k_g$ represents the difference
between energy loss at the glottal inlet and pressure
recovery at the outlet. This coefficient depends not only on the
glottal geometry but also on the glottal opening, subglottal pressure,
and flow through the glottis \cite{Fulcher:2011}. \RevisionText{It 
should be noted that equations~\eqref{eq:vout}--\eqref{eq:rtot} bear 
resemblance to the description of airflow in \cite{Ruty:2007,Ruty:2008} 
where the \SecondRevision{pressure loss and recovery} effects, however, 
are accounted for by flow separation in a diverging channel.}

The pressure $p=p(x,t)$ in the glottis is given in terms of $v_0$ from
equation~\eqref{eq:vout} by \SecondRevision{making use of the mass
  conservation} and the Bernoulli theorem $p(x,t)+\frac{1}{2} \rho
V(x,t)^2=p_{sub}$ for static flow. Since each vocal fold has two
degrees of freedom, the pressure $p$ in the glottis and the VT/SGT
counter pressure $p_c$ can be reduced to an aerodynamic force pair
$F_A = \bbm{F_{A,1} & F_{A,2}}^T$ where $F_{A,1}$ affects at the right
(i.e., the superior) end of the glottis ($x=L$) and $F_{A,2}$ the left
(i.e., the inferior) end ($x=0$) in Figure~\ref{fig:model}
(\SecondRevision{lower left} panel). This reduction can be carried out
by using the total force and moment balance equations
\begin{equation} \label{eq:int2} 
  \begin{aligned}
    F_{A,1} + F_{A,2} & = h \int_0^L (p(x,t)-p_{sub}) \, dx    \text{ and }  \\
    L \cdot F_{A,1} & = \frac{h}{\cos^2 \phi} \int_0^L x (p(x,t)-p_{sub}) \, dx-L C_{pc} p_c, 
  \end{aligned}
\end{equation}
\SecondRevision{where $\phi$ is the angle of the inclined vocal fold
  surface from the horizontal as shown in Figure~\ref{fig:model}
  (right panel), and $C_{pc}=C_{pc}(t)$ accounts for the moment arms
  and areas on which $p_c$ acts. An expression for $C_{pc}$ is given
  in equation~\eqref{eq:pc_coeff}. The force calculations are done
  using the pressure difference $p(x,t)-p_{sub}$ because we assume
  that $w_{ij} = 0$ for all $i,j = 1,2$ is the equilibrium under
  subglottal pressure $p_{sub}$ \ThirdRevision{(making the simplifying
    approximation that this equals the ambient hydrostatic pressure in
    the tissues surrounding the vocal folds)}, and therefore forces
  $F_{A,1}$ and $F_{A,2}$ must vanish when $p(x,t) \equiv p_{sub}$ and
  $p_c = 0$.}  \RevisionText{Note that \SecondRevision{since the
    displacements $w_i$ are in the y-direction only,} the aerodynamic
  forces have been assumed to act \SecondRevision{in this direction as
    well} as shown in Figure~\ref{fig:model}} \SecondRevision{(lower
  left panel)}.  The moment is evaluated with respect to point
$(x,y)=(0,0)$ for the lower fold and $(x,y)=(0,H_0)$ for the upper
fold in Figure~\ref{fig:model} (right panel).  Evaluation of these
integrals yields
\begin{equation} \label{eq:aeroforces}
\begin{aligned}
      F_{A,1} & =  \frac{\rho h L H_1^2 v_o^2}{2 \cos^2\phi} \left(-
        \frac{1}{\Delta W_1(\Delta W_2-\Delta
          W_1)} +\frac{1}{(\Delta W_1-\Delta W_2)^2}
        \ln{\left(\frac{\Delta W_2}{\Delta W_1} \right)}\right) \\
        & - C_{pc} p_c,  \quad  \mathrm{ for } \quad \Delta W_1> 0, \quad \mathrm{ and } \\
     F_{A,2} & =  \frac{\rho h L H_1^2 v_o^2}{2 \cos^2\phi}  \left(
        \frac{\sin^2\phi \Delta W_2+ \cos^2\phi \Delta W_1}{\Delta W_1\Delta W_2(\Delta W_2-\Delta
          W_1)}-\frac{1}{(\Delta W_1-\Delta W_2)^2}
        \ln{\left(\frac{\Delta W_2}{\Delta W_1} \right) }\right) \\
        & +  	 C_{pc} p_c, \quad  \mathrm{for} \quad \Delta W_1> 0.
\end{aligned}
\end{equation}
  During the glottal closed phase (i.e., when $\Delta W_1(t) <
  0$), the aerodynamic force in equations \eqref{eq:aeroforces} is not
  enabled\SecondRevision{, and the vocal fold load force is instead 
  given by equation \eqref{HertzImpactEq} above.}

\section{\label{VTASec} Vocal Tract and Subglottal Acoustics}

\subsection{Modelling VT acoustics by Webster's equation}

A generalised version of Webster's horn model resonator is used as
acoustic loads to represent both the VT and the SGT. It is given by
\begin{equation} \label{eq:webster}
	\frac{1}{c^2 \Sigma (s)^2} \frac{\partial ^2 \psi}{\partial t^2} + 
	\frac{2\pi \alpha_1 W(s)}{A(s)} \frac{\partial \psi}{\partial t} - 
	\frac{1}{A(s)} \frac{\partial}{\partial s} 
	\left(A(s) \frac{\partial \psi}{\partial s}\right) = 0,
\end{equation}
where $c$ denotes the speed of sound, the parameter $\alpha_1 \geq 0$
regulates the energy dissipation through air/tissue interface, and the
solution $\psi = \psi(s,t)$ is the velocity potential of the acoustic
field. Then the sound pressure is given by $p=\rho \psi_t$ where
$\rho$ denotes the density of air. The generalised Webster's model for
acoustic waveguides has been derived from the wave equation in a
tubular domain in \cite{Lukkari:WECD:2013}, its solvability and 
energy notions have been treated in \cite{Aalto:AWG:2015}, and 
the approximation properties in \cite{Lukkari:APostError:2015}.

The generalised Webster's equation \eqref{eq:webster} is applicable if
the VT is approximated as a curved tube of varying cross-sectional
area and length $L_{VT}$. The centreline $\gamma : [0, L_{VT}]
\longrightarrow \mathbb{R}^3 $ of the tube is parametrised using
distance $s \in [0, L_{VT}]$ from the superior end of the glottis, and
it is assumed to be a smooth planar curve. At every $s$, the
cross-sectional area of the tube perpendicular to the centreline is
given by the area function $A(s)$, and the (hydrodynamic) radius of
the tube, denoted by $R(s) > 0$, is defined by $A(s) = \pi R(s)^2$.
The curvature of the tube is defined as $\kappa (s) := \Vert \gamma''
(s) \Vert$, and the curvature ratio as $\eta (s) := R(s)\kappa
(s)$. Since the tube does not fold on to itself, we have always $\eta
(s) < 1$, and clearly $\eta \equiv 0$ if the tube is straight.

We are now ready to describe the rest of the parameters appearing in
equation~\eqref{eq:webster}: They are the stretching factor $W(s)$ and the
sound speed correction factor $\Sigma(s)$, defined by
\begin{equation}
\label{eq:webstercorr}
\begin{aligned}
	W(s) & :=  R(s)\sqrt{R'(s)^2 + (\eta(s)-1)^2}, \\
	\Sigma(s) & :=  \left(1+\tfrac{1}{4}\eta^2(s)\right)^{-1/2}.
\end{aligned}
\end{equation}
In the context of VT, we use the following boundary conditions 
for equation~\eqref{eq:webster}:
\begin{equation} \label{eq:boundaries}
	\left\lbrace \begin{array}{r c l}
	\frac{\partial \psi}{\partial t}(L_{VT},t) + \theta c \frac{\partial  \psi}
	{\partial s}(L_{VT},t) & = & 0, \\
	\frac{\partial \psi}{\partial s}(0,t) & = & -c_1 v_0(t).
	\end{array} \right.
\end{equation}
The first boundary condition is imposed at the mouth opening, and the
parameter $\theta \geq 0$ is the normalised acoustic resistance due to
exterior space.  \SecondRevision{The values for $\theta$ are based on
  the piston model given in \cite[Chapter~7,
    Eq.~(7.4.31)]{Morse:1968}. However, the acoustic impedance of the
  piston model has a significant reactive part as well, and its effect
  has been investigated by replacing the first equation in
  \eqref{eq:boundaries} by another boundary condition that corresponds
  to the impedance $Z(s) = \frac{s RL}{R + sL}$. \ThirdRevision{The
    rational impedance of the same form appears also as the
    ``first-order high pass model'' for termination of an acoustic
    horn in \cite[Section~4.1]{Helie:Radiation:2003}. In the present
    article,} the nominal values for $R$ and $L$ have been obtained by
  interpolation from the impedance of the piston model as given in
  Table~\ref{tbl2:VT} below where also the effects of the reactive
  component are discussed.}

The latter boundary condition in equation~\eqref{eq:boundaries}
couples the resonator to the glottal flow given by
equation~\eqref{eq:vout}. The scaling parameter $c_1 = h H_1 /A(0)$
extends the assumption of incompressibility from the control area just
right to the glottis in Figure~\ref{fig:model} (right panel) to the VT
area slice nearest to the glottis. \ThirdRevision{Using $c_1$ and a VT
  geometry independent control area, instead of defining $v_o$ as the
  flow through $A(0)$ directly, reduces the sensitivity of the model
  to the accurate placement of the glottis in the VT geometries which
  can be problematic in MRI data.}

\subsection{\label{sec:SGT} Subglottal tract acoustics}

Anatomically, the SGT consists of the airways below the larynx: trachea, 
bronchi, bronchioles, alveolar ducts, alveolar sacs, and alveoli. This 
system has been modelled either as a tree-like structure \cite{Ho:SGT:2011} or, more simply, as an acoustic horn whose area 
increases towards the lungs \cite{Birkholz:Turbulence:2007,Lous:1998}.  
We take the latter approach and denote the cross-sectional area and the
horn radius by $A_s(s)$ and $R_s(s)$, respectively, where
$s\in[0,L_{SGT}]$ and $L_{SGT}$ is the nominal length of the SGT.

Since the subglottal horn is assumed to be straight, i.e. $\eta
\equiv 0$, we have $\Sigma \equiv 1$ and $W_s(s) =
R_s(s)\sqrt{R_s'(s)^2 + 1}$.  Then
equations~\eqref{eq:webster}--\eqref{eq:boundaries} translate to
\begin{equation} \label{eq:boundaries_sg}
	\left\lbrace \begin{array}{r c l}
 	\frac{1}{c^2} \frac{\partial ^2 \widetilde{\psi}}{\partial t^2} + 
 	\frac{2\pi \alpha_2 W_s(s)}{A_s(s)} \frac{\partial \widetilde{\psi}}{\partial t} -
 	 \frac{1}{A_s(s)} \frac{\partial}{\partial s} \left(A_s(s) 
 	 \frac{\partial \widetilde{\psi}}{\partial s}\right) & = & 0,	\\
 	 \frac{\partial \widetilde{\psi}}{\partial t}(L_{SGT},t) + 
	\theta_s c \frac{\partial  \widetilde{\psi}}{\partial s}(L_{SGT},t) & = & 0, \\
	\frac{\partial \widetilde{\psi}}{\partial s}(0,t) & = & c_2 v_0(t),
	\end{array} \right.
\end{equation}
where the solution $\widetilde{\psi}$ is the velocity potential for
the SGT acoustics. We now use the scaling parameter value $c_2 = 
h H_1/A_s(0)$.  \RevisionText{The same kind of losses are considered in
  the SGT as in the VT: a termination loss characterised by normalised
  acoustic resistance $\theta_s \geq 0$, and energy dissipation
  through the air/tissue interface along the length of the horn regulated
  by parameter $\alpha_2 \geq 0$.}

\subsection{\label{CounterPressureSec} The acoustic counter pressure}

The final part of the vowel model produces the feedback coupling from
VT/SGT acoustics back to glottal oscillations.  This coupling is
realised by \SecondRevision{the product of} the acoustic counter 
pressure $p_c = p_c(t)$ \SecondRevision{and the coupling coefficient
$C_{pc} = C_{pc}(t)$ as already shown} in equations~\eqref{HertzImpactEq} and
\eqref{eq:aeroforces} above. 

The counter pressure is the resultant of sub- and supraglottal pressure 
components, and it is given in terms of velocity potentials from equations~\eqref{eq:webster} and
\eqref{eq:boundaries_sg} by
\begin{equation} \label{eq:pc}
  p_c(t) = Q_{pc} \rho \left ( \psi_t(0,t) - c_3
  \widetilde{\psi}_t(0,t) \right ).
\end{equation}

The tuning parameter $Q_{pc} \in [0,1]$ enables scaling the magnitude
of the feedback from the VT and SGT resonators to the vocal folds.
The parameter $Q_{pc}$ is necessary because it is difficult to
estimate from anatomic data the area on which the counter pressure
$p_c$ acts.  In simulations, excessive acoustic load forces lead to
non-stationary, even chaotic vibrations of the vocal folds.

The second parameter $c_3 \geq 0$ in equation~\eqref{eq:pc} accounts for
the differences in the areas and moment arms for the supra- and
subglottal pressures that load the equations of motion
equations~\eqref{eq:liikeyhtalot} for vocal folds.  Based on the idealised
vocal folds geometry in Figure~\ref{fig:model} \SecondRevision{(right panel)}, 
we obtain an overly high
nominal value $c_3=8.6$. In the simulations of this article, we use
$Q_{pc}$ as a tuning parameter to obtain the desired glottal pulse
waveform, and the value of $c_3$ is kept fixed (one could say,
arbitrarily) at $c_3 = 1$ (if the subglottal resonator is coupled)
or $c_3 = 0$ (if the subglottal acoustics is ignored). If it is
necessary for producing a realistic balance between supra- and
subglottal feedbacks, the value of $c_3$ can be increased without
losing stable phonation up to $Q_{pc}c_3 \approx 0.6$.

\SecondRevision{The coupling coefficient $C_{pc}$ is best understood
  in reference to the moment balance  in equation \eqref{eq:int2},
  although it appears in the same way in both equations
  \eqref{HertzImpactEq} and \eqref{eq:aeroforces}.  The counter
  pressure $p_c$ is assumed to affect only in the longitudinal
  direction (i.e., vertically in right panel of
  Figure~\ref{fig:model}). For each vocal fold, $p_c$ acts on an area
  of $\frac{H_1-\Delta W_1}{2}h$ and produces a moment arm of
  $\frac{2H_0-H_1-\Delta W_1}{4}$ around points $(x,y)=(0,0)$ and
  $(x,y)=(0,H_0)$ for the lower and upper folds, respectively. Hence}
\begin{equation}\label{eq:pc_coeff}
C_{pc} = \frac{(H_1-\Delta W_1)(2H_0-H_1-\Delta W_1)}{8L}.
\end{equation}

\section{Anatomic Data and Model Parameters} 

\begin{figure} [!t]
  \centering
\includegraphics[width=5cm]{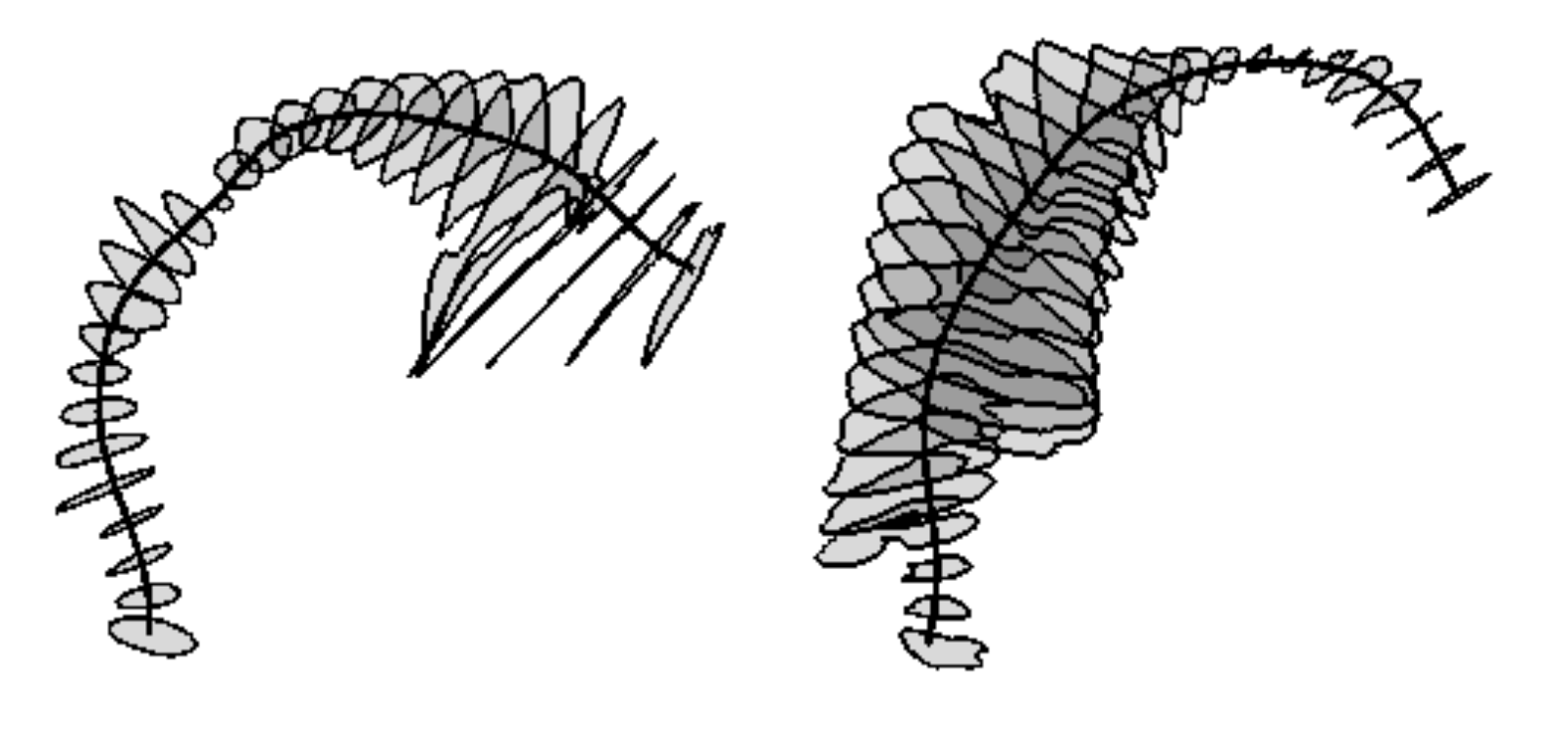} \hspace{2mm}
\includegraphics[width=6cm]{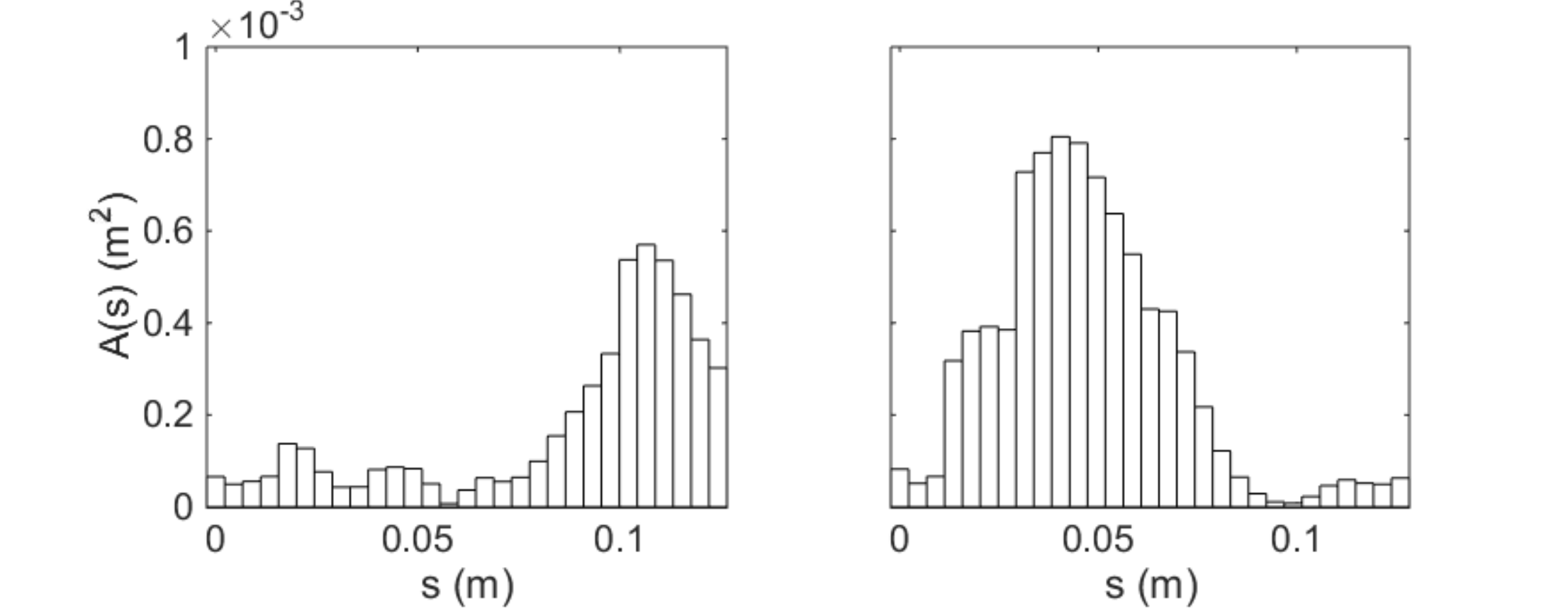} \\
\caption{Left: The VT intersections extracted from the test subject during phonation
of [\textipa{\textscripta}] and [\textipa{i}].  Right: The resulting
area functions for equation~\eqref{eq:webster}, represented as a function
of distance from the glottis.}
\label{VT-Area-fig}
\end{figure}

\subsection{Area functions for VT and SGT}

Solving Webster's equation requires that the VT is represented with an
area function and a centreline, from which curvature information can
be computed.  Two different VT geometries corresponding to vowels from
a healthy 26 years old female\footnote{In fact, she is one of test
  subjects in the experimental companion article
  \cite{A-M-V:MLBVFVTOPII}.} are used: A prolonged
\textipa{[\textscripta]} produced at fundamental frequency
$f_o=168$~Hz and similarly produced \textipa{[i]} at $f_o=210$~Hz.
These geometries have been obtained by Magnetic Resonance Imaging
(MRI) using the experimental setting that has been described in
\cite{Aalto:MesTec:2014}; see also
\cite{S-T-H:VTAFMRI,Story:FemaleMRI:1998,Story:VTparam:1998} for
earlier approaches. 
The extraction of the computational geometry from raw MRI data has
been carried out by the custom software described in
\cite{Kivela:MT:2015,Ojalammi:Segmentation:2017}.  The VT geometries and the area
functions are shown in Figure~\ref{VT-Area-fig}, and related VT
geometry dependent parameter values are given in Table~\ref{tbl:VT}.

The piston model \cite[Chapter~7]{Morse:1968} gives the expression
$\theta = 2 \pi A(L_{VT})/ \ell^2$ for the normalised acoustic
resistance in equation~\eqref{eq:boundaries} where we use the nominal
wavelength $\ell = 171.5 \, \mathrm{mm}$, \ThirdRevision{corresponding
  to the centre frequency $2 \, \mathrm{kHz}$ of the voice band}.  The
values for $\theta$[\textipa{\textscripta}], $\theta$[\textipa{i}]
together with the resonances $f_{Rj}$[\textipa{\textscripta}],
$f_{Rj}$[\textipa{i}] for $j = 1, 2$ are given in
Table~\ref{tbl:VT}. \SecondRevision{These values of the purely
  resistive load correspond to an infinitely long, non-resonant
  waveguide placed in front of the mouth, diameter of which is $95.2
  \, \textrm{mm}$ for [\textipa{\textscripta}] and $96.6 \,
  \textrm{mm}$ for [\textipa{i}].}

\begin{table}[ht] 
 \caption{\label{tbl:VT} Physical and physiological parameters dependent 
 on the VT geometry.}
 {\begin{tabular}{@{}lll@{}}  \toprule
  Parameter 	 & \textipa{[\textscripta]}  &  \textipa{[i]} \\   \midrule
 normalised acoustic resistance at mouth, $\theta$	 & \ThirdRevision{0.064} 
 			& \ThirdRevision{0.014}		\\
 area at mouth	 & 299 mm$^2$ &  66 mm$^2$ \\
 \ThirdRevision{VT} inertia parameter, $C_{iner}^{VT}$  & 2540 kg/m$^4$ & 2820 kg/m$^4$\\
 length of VT, $L_{VT}$	 &  132 mm  & 136 mm \\ 
 1st VT resonance, $f_{R1}$, from equations~\eqref{eq:webster}--\eqref{eq:boundaries} 		 	 		& 749 Hz 		& 199 Hz		 	\\ 
 2nd VT resonance, $f_{R2}$, from equations~\eqref{eq:webster}--\eqref{eq:boundaries}		 	 		& \ThirdRevision{2084} Hz 	& 2798 Hz	\\ \bottomrule
\end{tabular}} 
\end{table}

The MRI data that is used for the VT does not cover all of the SGT.
For this reason, an exponential horn is used as the subglottal area
function for equation~\eqref{eq:boundaries_sg}
\begin{equation} \label{eq:exp_horn}
   A_s(s) = A_s(0) e^{\epsilon s} \quad \text{ where } \quad \epsilon = \tfrac{1}{L_{SGT}}
   \ln{\left(\tfrac{A_s(L_{SGT})}{A_s(0)}\right)}
\end{equation}
following \cite{Birkholz:Turbulence:2007}. The values for $A_s(0) = 2
\, \mathrm{cm^2}$ and $A_s(L_{SGT}) = 10 \, \mathrm{cm^2}$ are taken
from Figure~1 in \cite{Birkholz:Turbulence:2007}. The horn length
$L_{SGT}$ is tuned so that the lowest subglottal resonance is $f'_{R1}
= 500 \, \mathrm{Hz}$ which results in the second lowest resonance at
$f'_{R2} = 1000 \, \mathrm{Hz}$.  This is a reasonable value for
$f_{R1}$ based on Table~1 in \cite{Austin:SGR:1997}; see also
\cite{Cranen:PM:1985,Cranen:SGF:1987}, \cite{Zanartu:AL1MM:2007} and
Figure 1 in \cite{Ho:SGT:2011}. \RevisionText{The SGT lung termination
  resistance in equation~\eqref{eq:boundaries_sg} is given the value
  $\theta_s = 1$ which corresponds to an absorbing boundary
  condition.} \ThirdRevision{The air column in this SGT model has a nominal
  inertia parameter value of 1040~kg/m$^4$ which is taken as a guideline for
  tuning the total inertia of the airways to obtain desirable flow pulse 
  waveforms. For the simulations in this article, $C_{iner}=1.5C_{iner}^{VT}$,
  where $C_{iner}^{VT}$ refers to the VT inertia parameters given in Table~\ref{tbl:VT}.}

\subsection{Static parameter values}

Table~\ref{tbl:consts} lists the numerical values of physiological and
physical constants used in all simulations. \ThirdRevision{Note that
the vocal fold springs are, for this study, placed symmetrically about
the midpoint of the vocal folds.} Based on the acoustic
reflection and transmission coefficients at the air/tissue interface,
the common value of the energy loss coefficients
\RevisionText{$\alpha_1$ and $\alpha_ 2$ in
  equations~\eqref{eq:webster} and \eqref{eq:boundaries_sg},
  respectively,} is taken as
 \begin{equation}  \label{eq:alpha}
 	\alpha_1 = \alpha_ 2 = \frac{\rho}{\rho_h c_h} =  7.6 \cdot 10^{-7} \, \mathrm{\frac{s}{m}}.
 \end{equation}

All the model parameter values introduced so far are assumed to be
equally valid for both female and male phonation, except for vocal fold
length $h$. As we are treating female phonation in this article, it
remains to describe the parameter values for
equations~\eqref{eq:liikeyhtalot} where the differences between female and
male phonation are most significant.  Hor{\'a}{\v c}ek \textit{et al.}  provide
parameter values for $M$ and $K$ for in male phonation 
\cite{Horacek:2005,Horacek:PhonThreshold:2002} but similar data for female
subjects cannot be found in literature. Instead, the masses in $M$ are
calculated by combining the vocal fold shape function used in
 \cite{Horacek:2005} with female vocal
fold length reported in \cite{Titze:MaleFemaleVoice:1989}. A first
estimate for the spring coefficients in $K$ is calculated by assuming
that the first eigenfrequency of the vocal folds matches the starting
frequency for the simulations.  The spring coefficients are then
adjusted until simulations produce the desired starting fundamental
frequency for the $f_o$-glides, giving the constant $K^0$ for
equations~\eqref{RisingScalingEq}--\eqref{FallingScalingEq}. For details of
these rather long calculations, see \cite{Aalto:MT:2009} and \cite{Murtola:MVP:2014}.

 \begin{table}[ht]
 \caption{\label{tbl:consts} Physical and physiological constants.}
{\begin{tabular}{@{}lll@{}}  \toprule
  Parameter 					 & Symbol	& Value \\ \midrule 
 speed of sound in air			 & $c$		& 343 m/s				\\
 density of air					 &$\rho$	& 1.2 kg/m$^3$				\\
 kinematic viscosity of air		 &$\mu$ 	& $18.27$ $\mu$N s/m$^2$		\\
 vocal fold tissue density 		 &$\rho_h$ 	& 1020 kg/m$^3$	 \\
 VT loss coefficient 		 	 &$\alpha_1$ 	& 7.6 $\cdot 10^{-7}$ s/m  \\
 SG loss coefficient 		 	 &$\alpha_2$ 	& 7.6 $\cdot 10^{-7}$ s/m  \\
 spring constant in contact (from \cite{Horacek:2005})  & $k_H$	& 730 N/m \\
 glottal gap at rest			 & $g$	& 0.3 mm	 		\\
 vocal fold length (from \cite{Titze:MaleFemaleVoice:1989})	& $h$	& 10 mm \\
 vocal fold thickness (from \cite{Horacek:2005})	 &$L$	& 6.8 mm     \\
\SecondRevision{superior vocal fold spring location (from \cite{Aalto:MT:2009})} & $l_1$ &\SecondRevision{0.85} \\
\SecondRevision{inferior vocal fold spring location (from \cite{Aalto:MT:2009})} & $l_2$ &\SecondRevision{0.15} \\
 control area height below glottis			 & $H_0$ 	& 11.3 mm \\
 control area height above glottis		 	 & $H_1$ 	& \SecondRevision{2 mm}	\\
 equivalent gap length for viscous loss in glottis & $L_g$ 	& 1.5 mm \\
 SGT length						 &$L_{SGT}$ & \SecondRevision{350} mm    \\ 
 normalised acoustic resistance at lungs 	 &$\theta_{s}$ & 1     \\ 
 glottal entrance/exit coefficient		 	 &$k_g$ 	& \SecondRevision{0.2} \\ 
 subglottal (lung) pressure over the ambient      & $p^0_{sub}$	& \SecondRevision{650	Pa} \\ \bottomrule \end{tabular}}
\end{table}

Let us conclude with a sanity check on the parameter magnitudes for
equation~\eqref{eq:liikeyhtalot} describing the vocal folds. The total
vibrating mass for female phonation is $m_{1}+m_{2}+m_{3}=0.27
\ \mathrm{g}$ and the total spring coefficients are $k_{1}+k_{2}=216
\ \mathrm{N/m}$. These nominal values yield $f_o \approx 150$ Hz for
female phonation. If the characteristic thickness of the vocal folds
is assumed to be about $5 \ \mathrm{mm}$, these parameters yield a
magnitude estimate for the elastic modulus of the vocal folds by $E
\approx \frac{k_{1}+k_{2}}{Lh} \cdot 5 \cdot 10^{-3} \ \mathrm{m}
\approx 15.9 \ \mathrm{kPa}$.  This should be compared to Figure~7 in
 \cite{Chhetri:Modulus:2011} where estimates are given for the
elastic modulus of \emph{ex vivo} male vocal folds and values
between $2.0 \ \mathrm{kPa}$ and $7.5 \ \mathrm{kPa}$ are proposed for
different parts of the vocal fold tissue.

\section{Computational Aspects}

\subsection{\label{ParConSec} Parameter control for obtaining vowel glides}

The $f_o$-glide is simulated by controlling two parameter values
dynamically.  First, the matrix $K$ is scaled while keeping the matrix
$M$ constant.  This approach is based on the assumption that the
vibrating mass of vocal folds is not significantly reduced when the
speaker's pitch increases; a reasonable assumption as far as register
changes are excluded. It should be noted that the relative magnitudes
of $M$ and $K$ essentially determine the resonance frequencies of
model \eqref{eq:liikeyhtalot}. However, attention must be paid to
their absolute magnitudes using, e.g., dimensional analysis since
otherwise the load terms $F_j(t)$ in equation~\eqref{eq:liikeyhtalot}
(containing the aerodynamic forces, contact force between the vocal
folds during the glottal closed phase, and the counter pressure from
the VT/SGT) would scale in an unrealistic manner.

The subglottal pressure, $p_{sub}$, is the second parameter used to
control the glide production. The dependence of the fundamental
frequency on $p_{sub}$ has been observed in simulations 
\cite{Ishizaka:1972,Sciamarella:2004}, \RevisionText{physical experiments using
upscaled replicas \cite{Ruty:2007}}, as well as in humans 
\cite{Lieberman:1969} and excised canine larynges \cite{Titze:Psub:1989}. 
The impact of $p_{sub}$ on $f_o$ is, however, secondary in these glides 
($f_o$ trajectories are within a few Hz). Instead, $p_{sub}$ is scaled in 
order to maintain phonation \RevisionText{and to prevent large changes in 
phonation type} 
as the stiffness of the vocal folds changes. \SecondRevision{The scaling 
parameter value of 2 was found by trial and error to maintain the glottal 
open quotient $OQ$  (proportion of glottal cycle during which the glottis 
is open),  the closing quotient $ClQ$ (proportion of the glottal cycle 
during which the flow is decreasing), and the maximum of $\Delta W_1$
approximately steady over the upward glide when acoustic feedback was 
disabled.}

The parameters are scaled exponentially with time
\begin{equation} \label{RisingScalingEq}
K(t) = 2.2^{2t/T} K^0,  \quad  p_{sub}(t) = 2^{t/T} p^0_{sub} 
\end{equation}
for rising glides, and 
\begin{equation} \label{FallingScalingEq}
K(t) = 2.2^{2-2t/T} K^0,  \quad  p_{sub}(t) = 2^{1-t/T} p^0_{sub} 
\end{equation}
for falling glides.  The duration of the glide is $T=3 \, \mathrm{s}$,
and $t$ is the time from the beginning of the
glide. \ThirdRevision{Note that the temporal scale of the glides is
  long compared to glottal cycles, and hence the control parameters
  $K$ and $p_{sub}$ can be regarded as static from the point of view
  of the vocal fold dynamics.} Other starting conditions
(particularly, vocal fold displacements and velocities, and pressure
and velocity distributions in the resonators) are taken from
stabilised simulations.  These parameters produce glides with $f_o$
approximately in the range $\GlideFrequencyBand$, although the exact
range depends on the VT geometry and vocal fold damping as well.

The damping parameters $b_{i}$ for $i=1,2,$ in equation~\eqref{eq:matrices}
play an important but problematic role in glottis models. If there is
too much damping (while keeping all other model parameters fixed),
sustained oscillations do not occur. Conversely, too low damping will
cause instability in simulated vocal fold oscillations. The magnitude
of physically realistic damping in vibrating tissues is not available,
and the present model could possibly fail to give a quasi-stationary
glottis signal even if realistic experimental damping values were
used.  With some parameter settings, the model even produces
quasi-stationary signal at several damping levels.  For simplicity, we
set $b_{i} = \beta > 0$ for $i=1,2$, and use golden section search to
find at least one value of vocal fold loss $\beta$ that results in
stable, sustained oscillation.  The damping remains always so small
that its lowering effect on the resonances of the mass-spring system
\eqref{eq:liikeyhtalot} is negligible.

\begin{figure} [!t]
  \centering
\hspace{1mm}
\includegraphics[width=\textwidth]{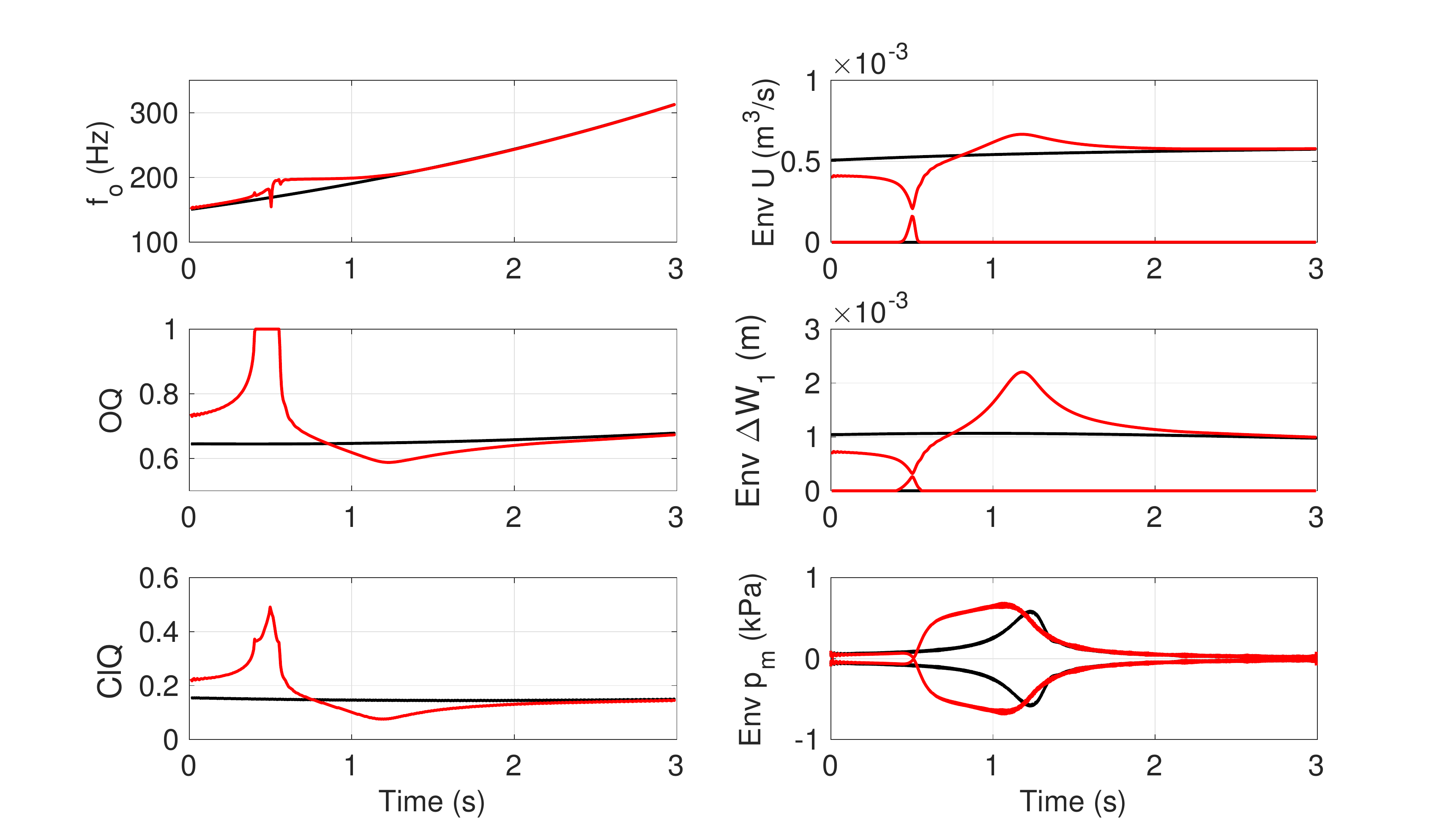}  				
\caption{\SecondRevision{Glide for vowel [\textipa{i}] with 
$Q_{pc} = 0.1$ and $\beta = 0.012 \, \textrm{kg/s}$ (red) 
and the same glide without VT and SGT feedback ($Q_{pc} = 0$) 
(black). Left: fundamental frequency ($f_o$), open quotient 
($OQ$), and closing quotient ($ClQ$). Right: \ThirdRevision{
Envelopes of volume flow ($U$), glottal opening ($\Delta W_1$), and sound pressure at lips ($p_{m}$)}. The values of $f_o$, $OQ$, and 
$ClQ$ have been extracted pulse by pulse from the volume 
flow signal.}}
\label{env_i}
\end{figure}

\subsection{Numerical realisation}

The model equations are solved numerically using MATLAB software and
custom-made code. The vocal fold equations of motion
\eqref{eq:liikeyhtalot} are solved by the fourth order Runge--Kutta
time discretisation scheme.  The flow equation \eqref{eq:vout} is
solved by the backward Euler method.  The VT \ThirdRevision{and SGT} are 
discretised by the FEM using piecewise linear elements (\ThirdRevision{$N=29$ for
VT and $N=10$ for SGT}) and the physical energy norm
of Webster's equation. Energy preserving Crank--Nicolson time
discretisation (i.e., Tustin's method \cite{Havu:CTT:2007}) is used. The
time step is almost always $10 \ { \mu s}$ which is small enough to
keep the frequency warping in Tustin's method under one semitone for
frequencies under $13 \mathrm{kHz}$. Reduced time step, however, is
used near glottal closure.  This is due to the discontinuity in the
aerodynamic force in equation~\eqref{eq:aeroforces} at the closure which
requires numerical treatment by interpolation and time step reduction
as explained in Section~2.4.1 of \cite{Aalto:MT:2009}.

\ThirdRevision{Solving the equations of motion of the vocal folds is
  the computationally most expensive part of the model, taking
  approximately 55\% of the running time in simulations of steady
  phonation with given parameter values. In comparison, solving the
  Webster's equations with precomputed mass, stiffness, and loss
  matrices takes approximately 10\% of the simulation time, and the
  flow equation solver less than 2\%.}

\begin{figure} [!t]
  \centering
\hspace{1mm}
\includegraphics[width=\textwidth]{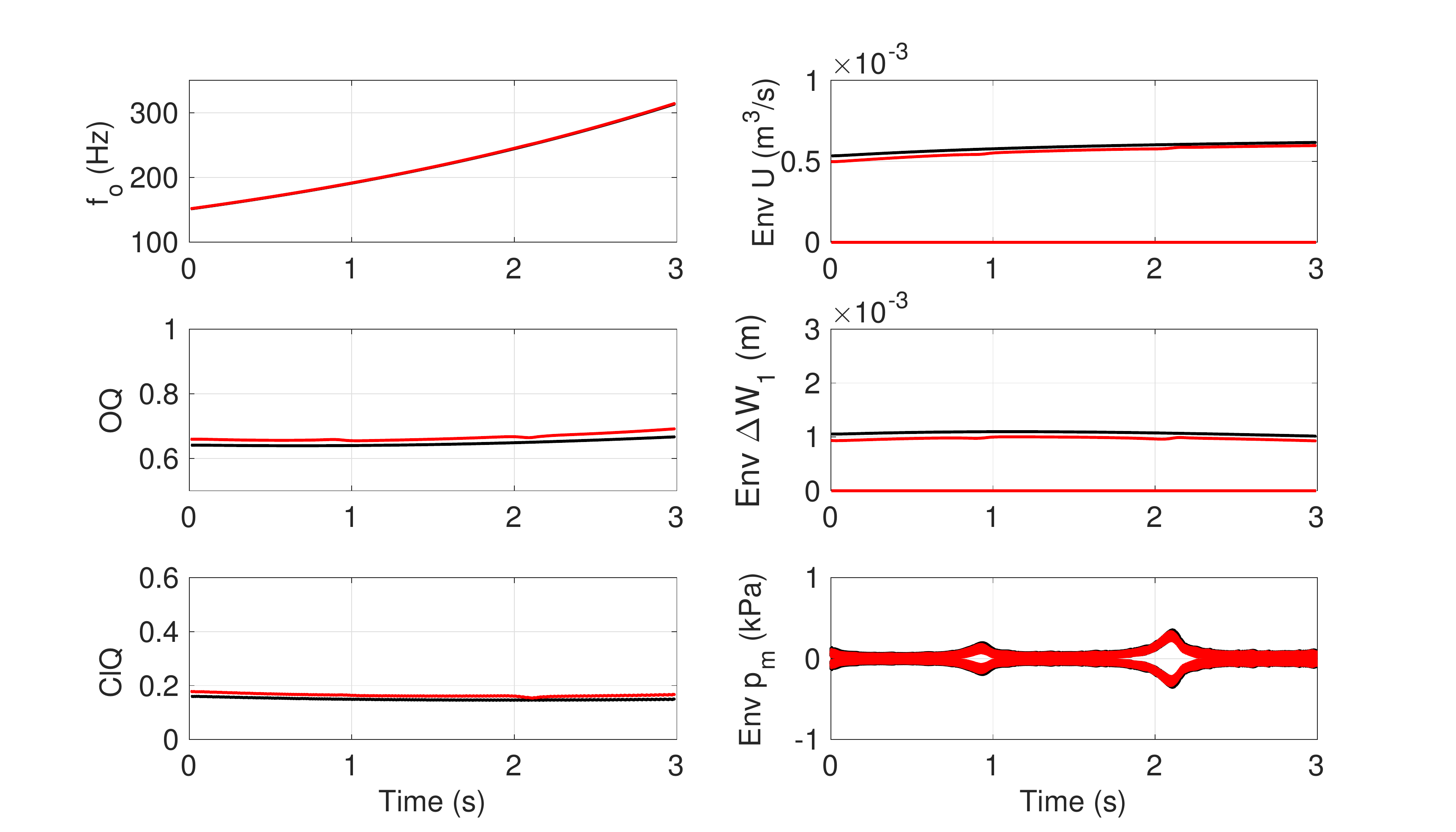}  					
\caption{\SecondRevision{Glide for vowel [\textipa{\textscripta}] with 
$Q_{pc} = 0.1$ and $\beta = 0.012 \, \textrm{kg/s}$ (red)
and the same glide without VT and SGT feedback ($Q_{pc} = 0$) 
(black). Left: fundamental frequency ($f_o$), open quotient 
($OQ$), and closing quotient ($ClQ$). Right: \ThirdRevision{
Envelopes of volume flow ($U$), glottal opening ($\Delta W_1$), 
and sound pressure at lips ($p_{m}$)}. $f_o$, $OQ$, and 
$ClQ$ have been extracted pulse by pulse from the volume 
flow signal.}}
\label{env_a}
\end{figure}

\section{\label{SimSec} Simulation Results} 

The results of \SecondRevision{upward} glide simulations for vowels 
[\textipa{\textscripta, i}] are shown in Figures~\ref{env_i}--\ref{env_a}.  
The fundamental frequency $f_o$ trajectory \SecondRevision{as well as 
glottal open quotient $OQ$ and closing quotient $ClQ$ have} been extracted 
from the glottal volume flow $U$ signal \SecondRevision{pulse by 
pulse} in all figures. \ThirdRevision{Envelopes of $U$, glottal 
opening $\Delta W_1$, and pressure signal at lips $p_m$ are also displayed.}

The simulations indicate a consistent locking pattern at
$f_{R1}[\textipa{i}]$ in $f_o$ trajectories that vanishes if the VT
feedback is decoupled by setting $Q_{pc} = 0$. The locking pattern in
rising glides follows the representation given in Figure~\ref{qual_b_Q}
(right panel): sudden jump upwards to $f_{R1}$, a locking to a plateau
level, and a smooth release.  Such locking behaviour is not observed
for glides of [\textipa{\textscripta}] where
$f_{R1}$[\textipa{\textscripta}] is not inside the simulated frequency
range $\GlideFrequencyBand$. The \SecondRevision{vocal tract} 
\ThirdRevision{resonance fractions}
$f_{R1}$[\textipa{\textscripta}]$/4 = 187 \, \mathrm{Hz}$ and
$f_{R1}$[\textipa{\textscripta}]$/3 = 250 \, \mathrm{Hz}$, are within the
frequency range, and the corresponding events are visible in the sound
pressure signal at the lips; see Figure~\ref{env_a}.   They do not,
however, cause noticeable changes in the $f_o$ trajectory of the
glottal flow.

\SecondRevision{The frequency jump at $f_{R1}[\textipa{i}]$ in the
  simulations is preceded by a decrease in vocal fold
  oscillation and glottal flow amplitudes. This is accompanied by the
  disappearance of full glottal closure ($OQ=1$) and less sharp
  decrease in glottal flow during closure (higher $ClQ$), both of
  which indicate increased breathiness of the phonation. The locking
  plateau coincides with a nearly constant rate of decreasing $OQ$ and
  $ClQ$, and after the release of $f_o$ the parameters return to the 
feedback free trajectories.}

\begin{figure} [!t]
\centering
\hspace{1mm}
\includegraphics[width=\textwidth]{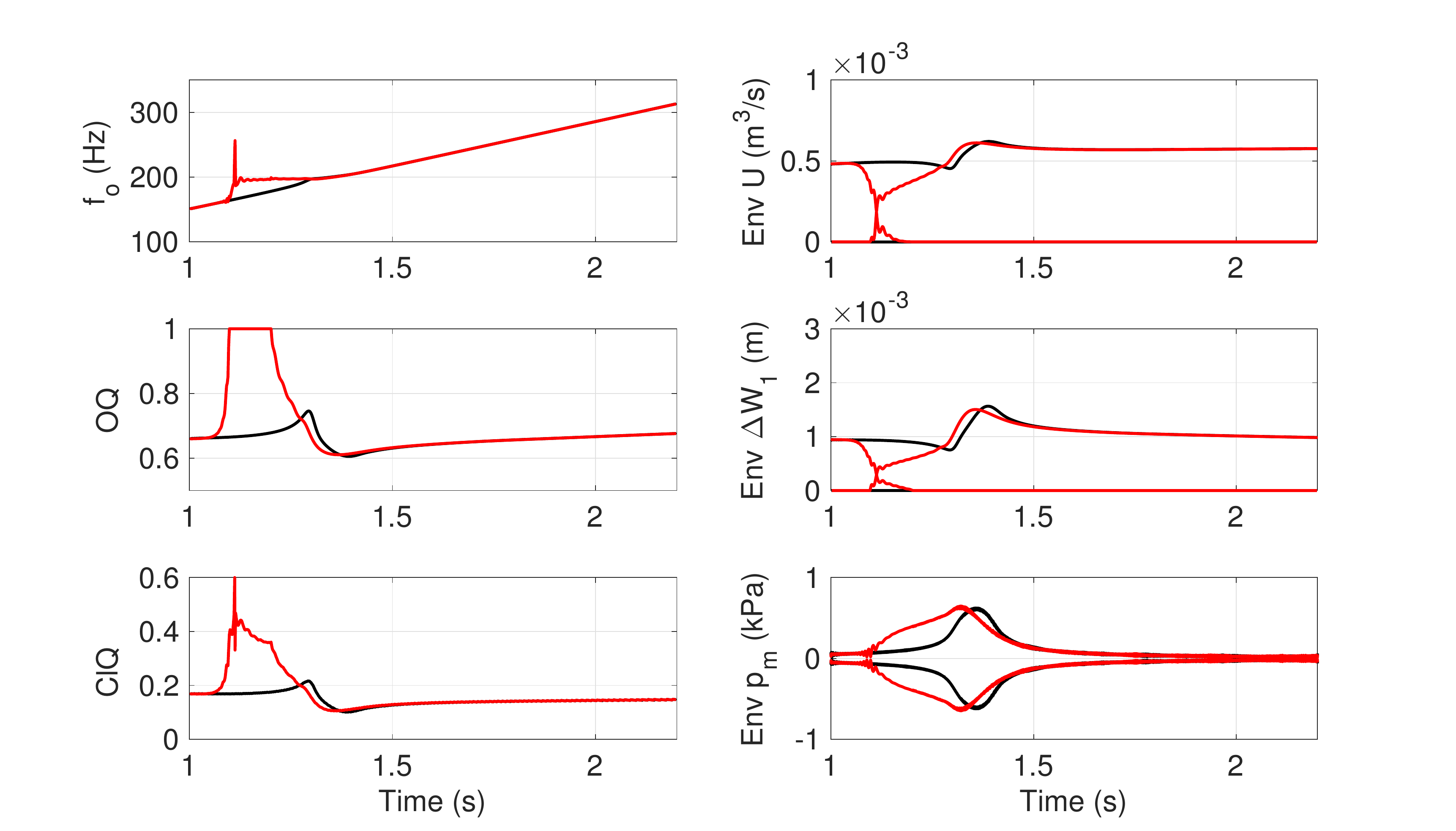}  				
\caption{\SecondRevision{Upward (black) and downward 
(red) glides for vowel [\textipa{i}] with $Q_{pc} = 0.04$ 
and $\beta = 0.012\, \textrm{kg/s}$. Left: fundamental 
frequency ($f_o$), open quotient ($OQ$), and closing quotient 
($ClQ$). Right: \ThirdRevision{Envelopes of volume flow ($U$), 
glottal opening ($\Delta W_1$), and sound pressure at lips 
($p_{m}$)}. On the 
x-axis, relative vocal fold stiffness refers to the coefficient 
of the $K^0$ matrix in equations \eqref{RisingScalingEq} and 
\eqref{FallingScalingEq}.}}
\label{env_updown}
\end{figure}

Keeping $Q_{pc}$ and other model parameters the same, a falling $f_o$
glide shows a \SecondRevision{significantly} more pronounced or longer 
locking at $f_{R1}$ compared to rising glides; see Figure~\ref{env_updown}. 
\SecondRevision{Note that in Figure~\ref{env_updown} the x-axis is the 
relative vocal fold stiffness, which for 
rising glides is $2^{t/T}$ and for falling glides $2^{1-t/T}$ as given 
in equations \eqref{RisingScalingEq} and \eqref{FallingScalingEq}.} 
\SecondRevision{The fluctuations in $f_o$ in the falling glides around 
the "corner" of the lock and at frequencies below this are qualitatively 
similar to what occurs at extreme values of $Q_{pc}$ and $\beta$ for 
rising glides. In contrast, fluctuations in the lip pressure envelope 
occur temporally after the release of the locking in both rising and 
falling glides.}

Finally, the effect of model parameters $\beta$ and $Q_{pc}$ on the glide 
simulations at $f_{R1}$[\textipa{i}] is considered. These observations are 
qualitatively described in Figure~\ref{qual_b_Q}. In the right panel, the 
medium values for $\beta$ refer to the interval $[0.01, 0.02]$ and for 
$Q_{pc}$ to the interval $[0.05, 0.1]$. These intervals can thus be regarded as
feasible parameter ranges for vowel glide simulations of [\textipa{i}].

Referring to Figure~\ref{qual_b_Q} (right panel), the full frequency
range $\GlideFrequencyBand$ for $f_o$ can be obtained with modal
locking as shown in Figure~\ref{env_i} if \SecondRevision{both $Q_{pc}$ and
$\beta$ have medium values or if one is high and the other low. If both 
parameters are high or one is high and other medium,} the simulated $f_o$ 
range \SecondRevision{is reduced} to above $200 \, \mathrm{Hz}$ which is the 
value of $f_{R1}$[\textipa{i}]. This glide starting frequency cannot be
lowered by changing $K$, and it appears to represent very strong modal
locking at the onset of the vowel glide simulation. 

The stability of glide simulations (understood as slowly changing
amplitude envelope of glottal volume flow $U$) becomes a serious issue
at low and high values of $\beta$.  We have tuned the subglottal
pressure $p_{sub}$ in glide simulations as given in
equations~\eqref{RisingScalingEq}--\eqref{FallingScalingEq}. 
\SecondRevision{If $p_{sub}$ were instead kept constant, we would observe
an increasing $OQ$ and decreasing amplitudes of glottal flow and vocal 
fold oscillations throughout the glide} but the qualitative behaviour of 
modal locking events\SecondRevision{, including the behaviour of phonation 
type parameters around these events,} remains very similar.

\section{\label{SensRob} Sensitivity and Robustness}

\begin{figure} [!t]
  \centering  			
 \includegraphics[scale=0.25]{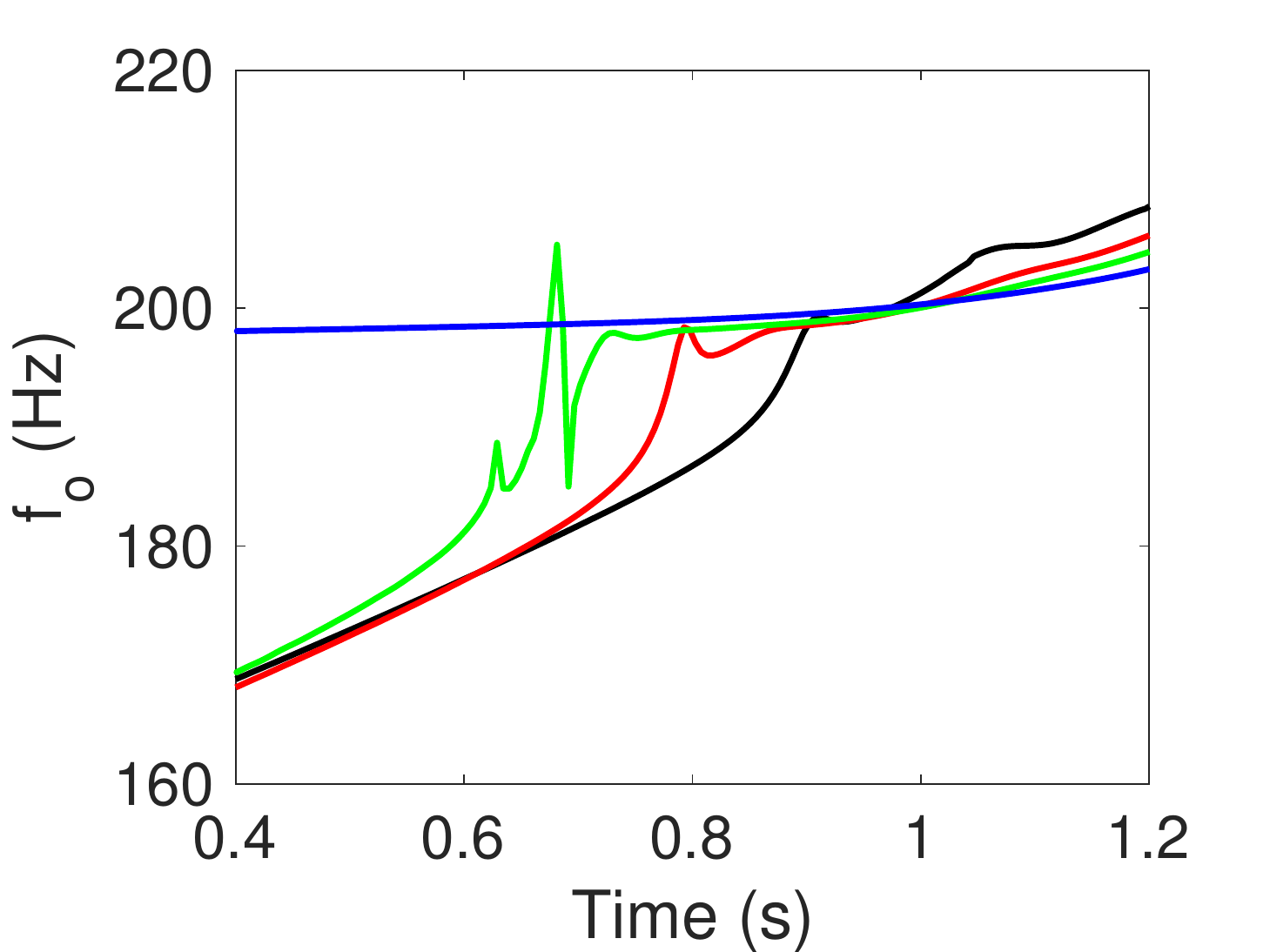}  			
 \includegraphics[scale=0.25]{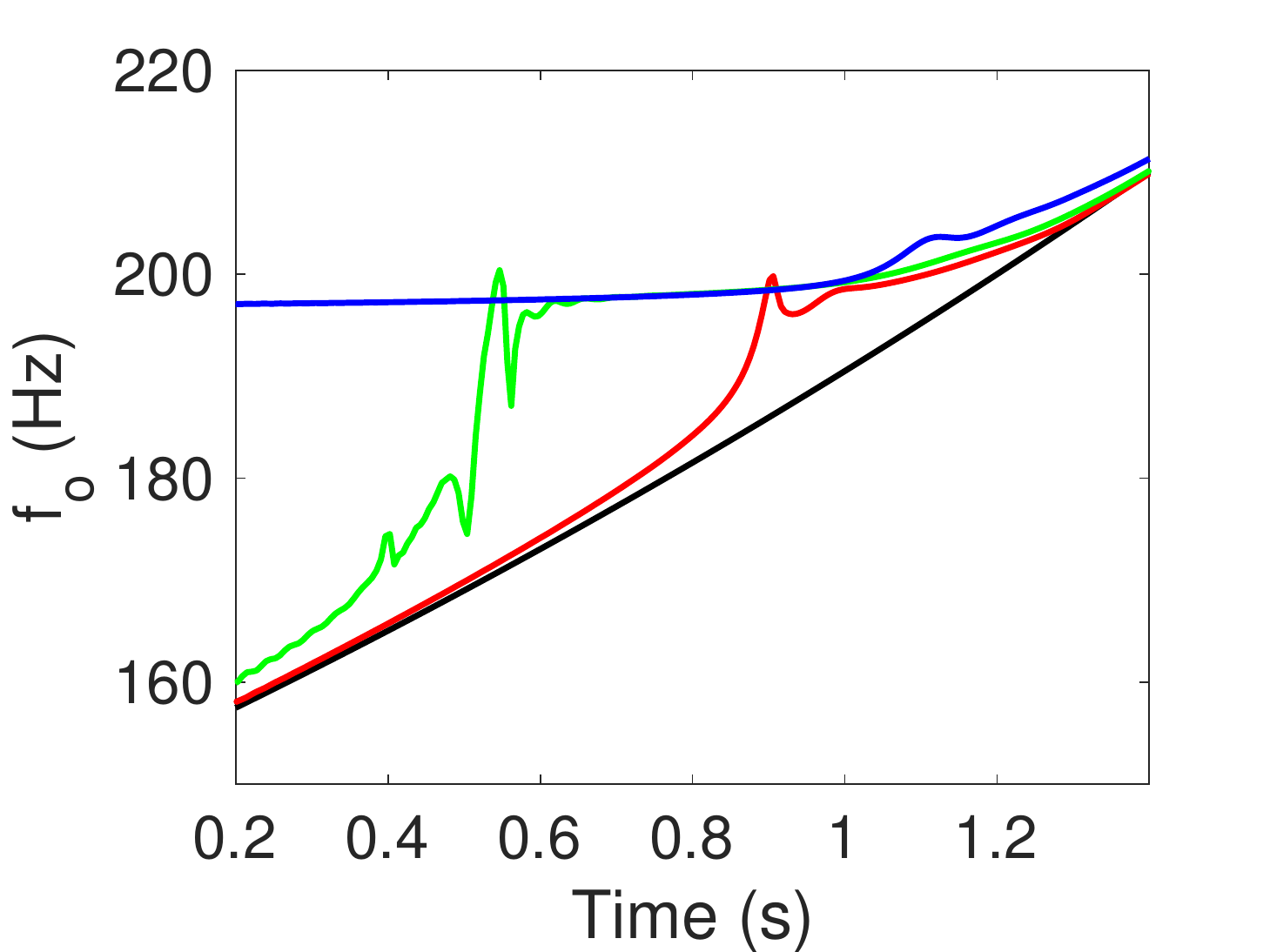}
 \includegraphics[scale=0.26]{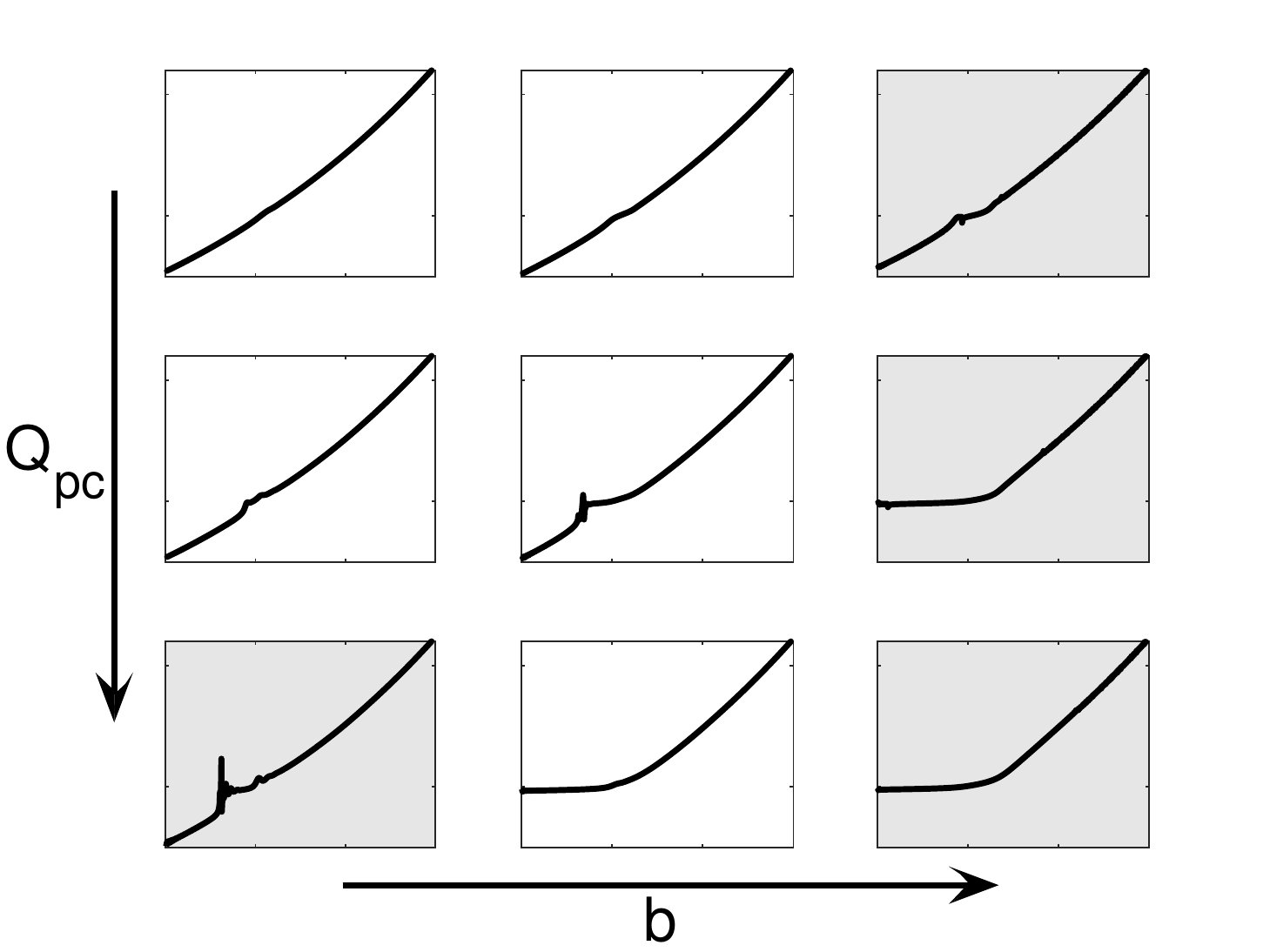}
\caption{Left: $f_o$ trajectories for [\textipa{i}] 
\SecondRevision{fixed $Q_{pc} = 0.1$ and different values of $\beta$: black 
  $0.005 \, \textrm{kg/s}$, red $0.01 \, \textrm{kg/s}$, green $ 0.015 \, 
  \textrm{kg/s}$, and blue $0.03 \, \textrm{kg/s}$.  
  Middle: $f_o$ trajectories for [\textipa{i}] with 
  $\beta = 0.012\, \textrm{kg/s}$ and different values of $Q_{pc}$: black 
  $0.0$, red $0.05$, green $ 0.1$, and blue $0.25$.
 Right: $f_o$ trajectories for [\textipa{i}] qualitatively as
 $Q_{pc}$ and $\beta$ increase in the direction of the arrow. Light
   gray background indicates that small parameter changes can lead 
   to loss of quasi-stable glides.}
}
\label{qual_b_Q}
\end{figure}

Parameter tuning of the vowel model is tricky business as can be seen
from model parameter optimisation experiments described in Chapter~4 in 
 \cite{Murtola:MVP:2014}. By exaggerating some of the parameter
values, it is possible to make vowel glide simulations over
$f_{R1}$[\textipa{i}] behave in a way that can be excluded by experiments
or observations from natural speech. 

In phonetically relevant simulations, various tuning parameters must
be kept in values that are not only physically reasonable but also do
not produce obviously counterfactual predictions. When such a
realistic operating point has been found, it remains to make sure that
the simulations give consistent and robust results near it. In doing
so, we also check which parts of the full model are truly significant
for the model behaviour reported above.

\subsection{Acoustics of the vocal tract by Webster's equation}

The \RevisionText{constants $\alpha_1$ and $\alpha_2$ in respective}
equations~\eqref{eq:webster} and \eqref{eq:boundaries_sg} regulate
the boundary dissipation at the air/tissue interface. As shown in
Section~3 in \cite{Lukkari:WECD:2013}, the same parameter appears in
the corresponding dissipating boundary condition $\alpha \phi_t + \bar
\nu \cdot \nabla \phi = 0$ for the wave equation $\phi_{tt} = c^2
\Delta \phi$ where $\phi$ is the 3D acoustic velocity potential and
$\bar \nu$ denotes the exterior normal of the VT/air boundary.  The
qualitative effect of physically realistic tissue losses to vowel
glide simulations was observed to be insignificant; see also Section~5
in \cite{Doel:Webster:2008}. However, these losses move slightly the
VT resonance positions computed from equations~\eqref{eq:webster}.

On the other hand, the VT resonances are quite sensitive to the normalised 
acoustic resistance $\theta$ in equation~\eqref{eq:boundaries}. This parameter 
regulates the energy loss through mouth to the external acoustic space, and its 
extreme values $0$ and $\infty$ correspond to open and closed ends for idealised
acoustic waveguides, respectively. Again, physically realistic variation in 
$\theta$ does not change the qualitative behaviour of vowel glides near
$f_{R1}$[\textipa{i}] as reported above.

\begin{figure} [!t]
  \centering
\includegraphics[width=0.3\textwidth]{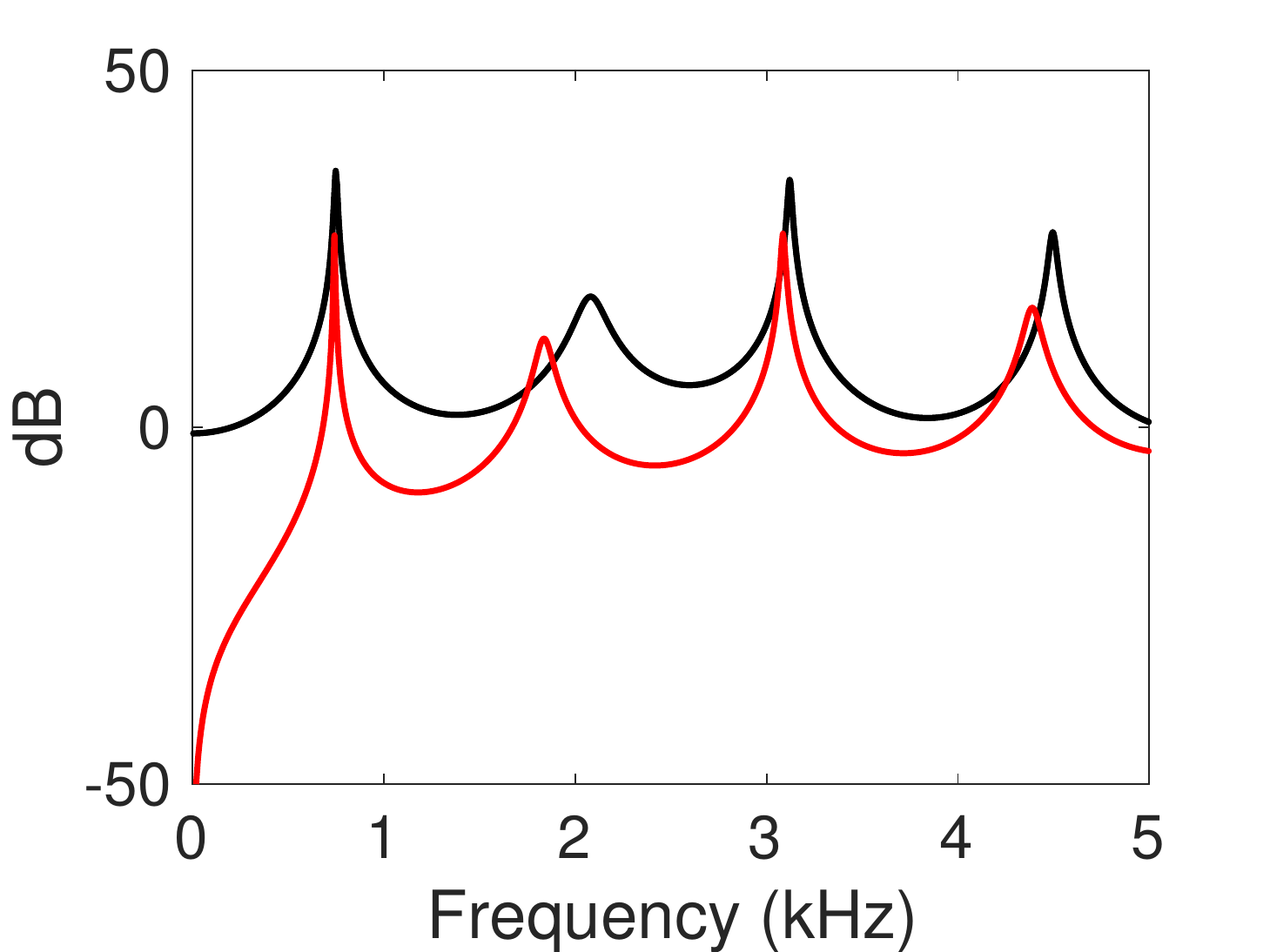} \hspace{2mm}
\includegraphics[width=0.3\textwidth]{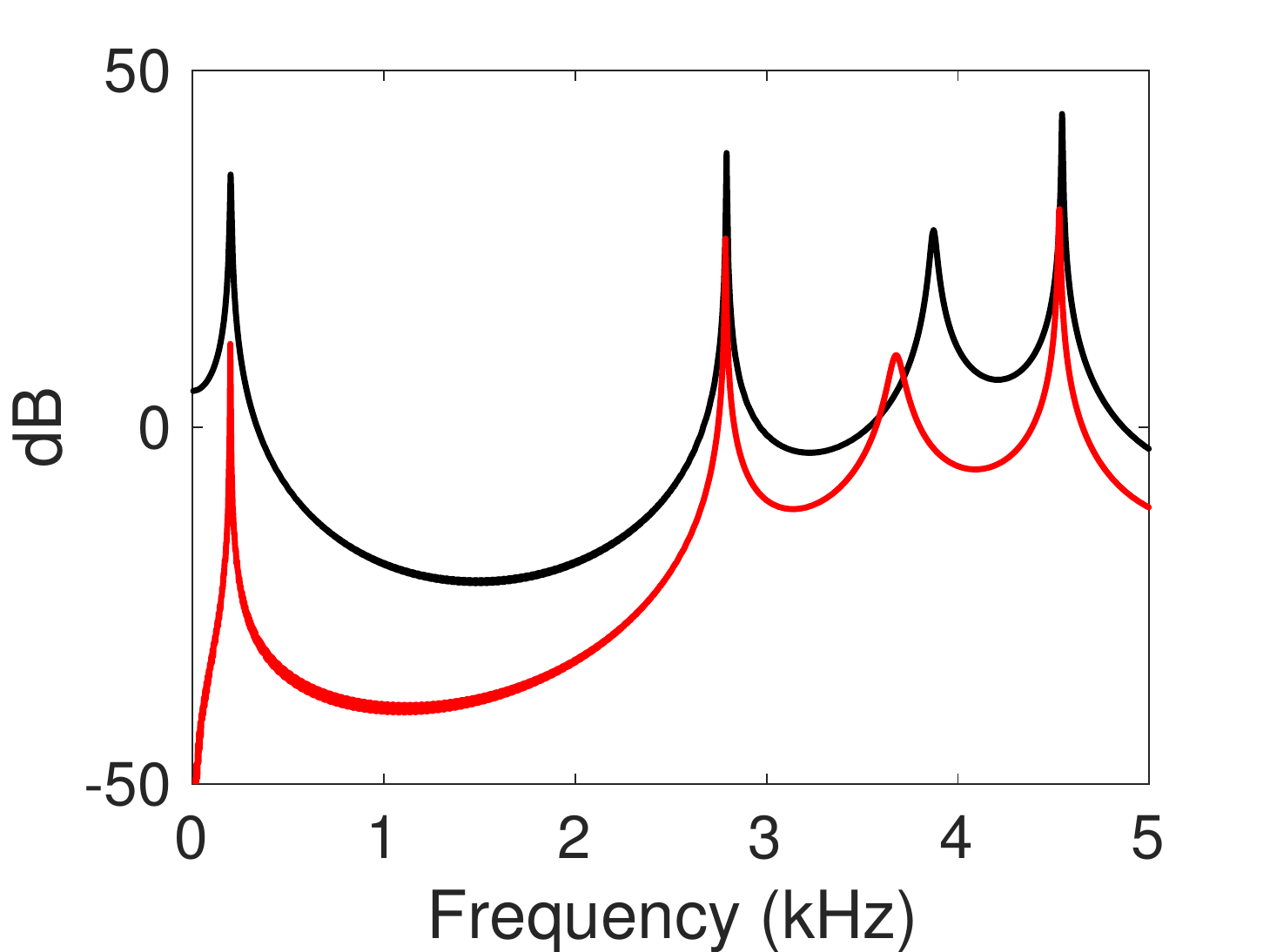} \hspace{2mm}
\includegraphics[width=0.3\textwidth]{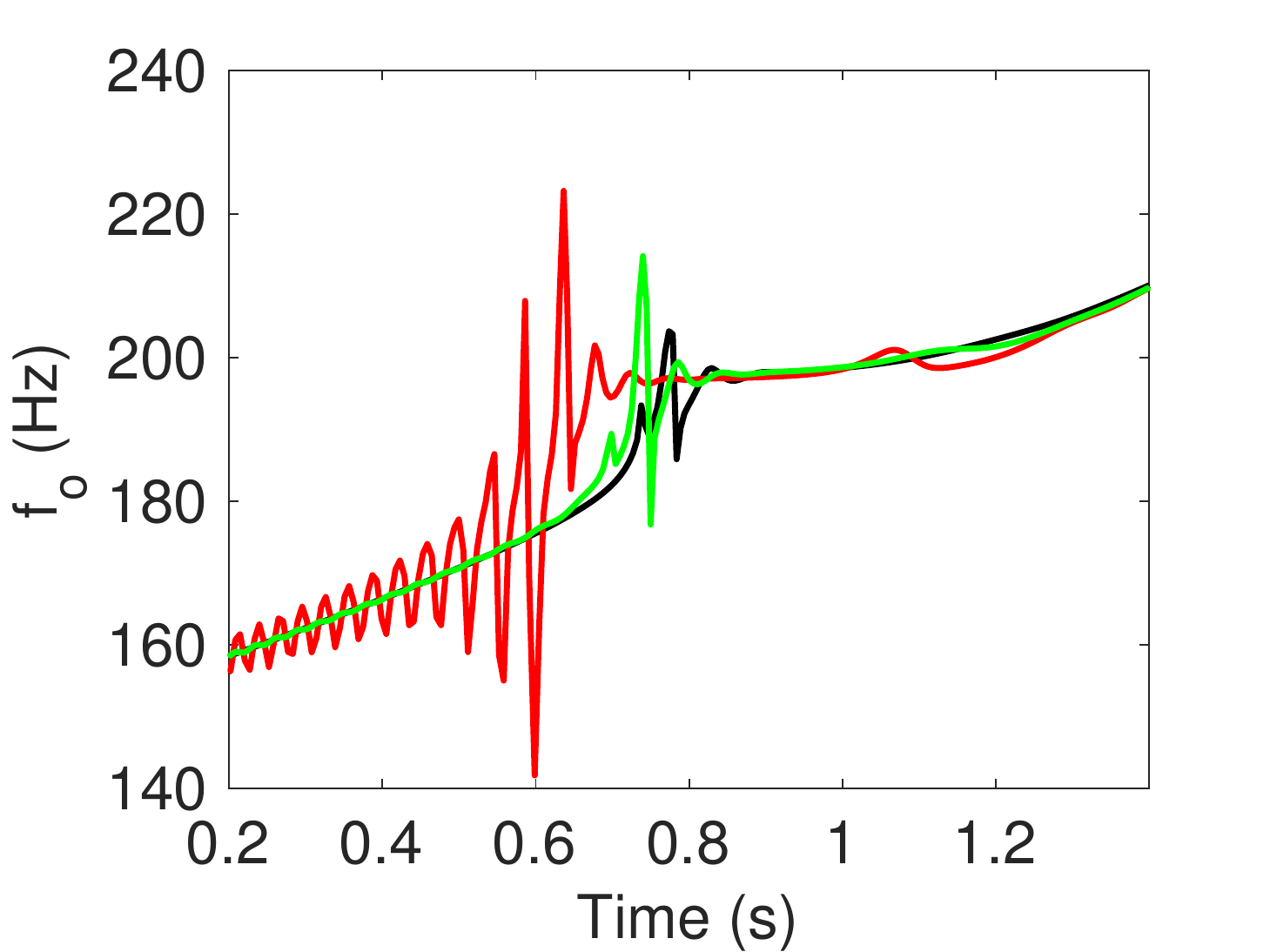} \\
\caption{ \SecondRevision{Left and middle: The frequency responses of
    the VT acoustic loads for \textipa{[\textscripta]} and
    \textipa{[i]}, computed from Webster's
    equation~\eqref{eq:webster}.  The black curves are the responses
    using the purely resistive load of equation~\eqref{eq:boundaries}
    at the mouth where the parameter values are given in
    Table~\ref{tbl:VT}. The responses using the RL impedance model and
    its nominal parameter values in Table~\ref{tbl2:VT} are shown in
    red.  Right: Ringing at a modal locking event of \textipa{[i]}
    shown for both types of loads at the mouth: purely resistive load 
    (black) and RL model load using nominal and tuned parameter values 
    (red and green, respectively). The parameter values
    for the RL impedance model are given in Table~\ref{tbl2:VT}.}}
\label{Effect-Reactive-Load}
\end{figure}

\SecondRevision{To consider the effect of the reactive acoustic
  loading at the mouth opening, a first order impedance model was used,
  based on a parallel coupling of a resistive load $R$ and an
  inductive load $L$.  The
  nominal values for $R$ and $L$ in Table~\ref{tbl2:VT} were obtained
  by interpolation at $200 \, \textrm{Hz}$ from the piston model given
  in \cite[Chapter~7, Eq.~(7.4.31)]{Morse:1968}. The transfer
  function $Z(s) = \frac{sRL}{R + sL}$ approximates the irrational
  piston model impedance very well for frequencies under $2 \,
  \textrm{kHz}$, and the frequency responses in
  Figure~\ref{Effect-Reactive-Load} (left and middle panels) are
  reasonable as well.}

\begin{table}[ht] 
 \caption{\label{tbl2:VT} \ThirdRevision{Values for the parameters of the RL 
 impedance model and its transfer function.}}
 {\begin{tabular}{@{}lll@{}}  \toprule
  Parameter 	 & \textipa{[\textscripta]}  &  \textipa{[i]} \\   \midrule
  Nominal value of $R$	 & $1.98 \cdot 10^6 \frac{\textrm{kg}}{\textrm{s m}^4}$	& $8.96 \cdot 10^6 \frac{\textrm{kg}}{\textrm{s m}^4}$ \\
  Nominal value of $L$	 & $33.2 \frac{\textrm{kg}}{\textrm{m}^4}$	& $70.6 \frac{\textrm{kg}}{\textrm{m}^4}$	\\
  Tuned value of $R$ & (Not required.) &  $8.96 \cdot 10^4 \frac{\textrm{kg}}{\textrm{s m}^4}$  \\ 
  \ThirdRevision{Nominal value of} $Z(400\pi i)$ & $879 + 4.17 \cdot 10^4 i$ & $879 + 8.87\cdot 10^4i$
  \\ \bottomrule
\end{tabular}} 
\end{table}

\SecondRevision{However, the value $\lim_{s \to +\infty}{Z(s)} = R$
  overestimates its piston model counterpart by over 700 \%, and the
  vowel simulations show excessive ringing, e.g., at modal locking
  events as shown in Figure~\ref{Effect-Reactive-Load} (right
  panel). In a low-order rational model, all of the lumped inductance
  appears at the mouth opening whereas the inductance is distributed
  by resistive shunting and transmission delays to an infinite volume
  in the piston model. Hence, the value of $R$ must be tuned down from
  its nominal value so as not to contradict experimental
  evidence.} 

  Another low-order \ThirdRevision{time-domain model for termination},
  based on an idealised spherical interface at a horn opening, is
  proposed in \cite[Eq.~(39)]{Helie:Radiation:2003}. \ThirdRevision{In 
  its most general form}, the model is an integro-differential delay equation
  with nine parameters and a single delay lag. Unfortunately, 
  \ThirdRevision{the general form} cannot be introduced to Webster's model 
  as a boundary  condition: this is the salient feature of the parallel RL model
  \ThirdRevision{(having the same circuit topology as the first-order high 
  pass model \cite[Eq.~(28)]{Helie:Radiation:2003})} that simplifies the
  implementation of the FEM solver.

The role of the VT curvature in equation~\eqref{eq:webster} is
involved, too. As can be seen from equation~\eqref{eq:webstercorr},
the curvature results in a second order correction in the curvature
ratio $\eta$ to the speed of sound $c$ in
equation~\eqref{eq:webster}.\footnote{It should be pointed out that
  equations~\eqref{eq:webster}--\eqref{eq:webstercorr} with
  nonvanishing $\eta$ is the ``right'' generalisation of Webster's
  horn model, corresponding to the wave equation in curved acoustic
  wave\-guides. This approach results in the approximation error
  analysis given in \cite{Lukkari:APostError:2015}. Somewhat
  paradoxically, a similar error analysis for the simpler model
  equations~\eqref{eq:webster}--\eqref{eq:webstercorr} with $\eta
  \equiv 0$ would require more complicated error estimation.} In
waveguides of significant intersectional diameter compared to
wavelengths of interest, the contribution of $\eta$ in
equation~\eqref{eq:webstercorr} appears to be secondary to a larger
error source that is related to curvature as well. This is caused by
the fact that a longitudinal acoustic wavefront does not propagate in
the direction of the geometric centreline of a curved waveguide even
if the waveguide were of circular intersection with constant diameter.
The wavefront has a tendency to ``cut the corners'' in a frequency and
geometry dependent manner, and we do not have a mathematically
satisfying description of the ``acoustically correct'' centreline that
would deal with this phenomenon optimally in the context of Webster's
equation.  Extraction of the area function $A(\cdot)$ for
equation~\eqref{eq:webster} from MR images, however, requires some
notion of a centreline, and using a different centreline would lead to
slightly different version of $A(\cdot)$. This would somewhat change,
e.g., the resonance frequencies of equation~\eqref{eq:webster} but not
the mathematical structure of the model nor the results of vowel glide
simulations.
Hence, we simply use the area functions and centrelines obtained from
3D MR images by the custom code described in
\cite{Ojalammi:Segmentation:2017}
using its nominal settings.

It remains to consider the non-longitudinal resonances of the VT.  By
its construction, the generalised Webster's equation does not take
into account at all the transversal acoustic dynamics of the VT. It is
known from numerical 3D Helmholtz resonance experiments on several
dozens of VT geometries that lowest non-longitudinal resonances of the
human VT tract are at approximately $4 \, \mathrm{kHz}$ corresponding to
$\lambda/2 \approx 4\, \mathrm{cm}$; see, e.g., \cite{Aalto:MesTec:2014} 
and \cite{Kivela:MT:2015}. Anatomically, such length 
may appear between opposing valleculae, piriform fossae, or even across 
the mouth cavity in some VT vowel configurations.  However, the upper 
limit of $4 \, \mathrm{kHz}$ for Webster's equation is adequate for the
computation of the acoustic counter pressure $p_c$ in
equation~\eqref{eq:pc} for several octaves lower fundamental frequencies
$f_o \in \GlideFrequencyBand$ that are used in vowel glide simulations.

\subsection{Subglottal acoustics}

To large extent, what was stated above about the modelling error of
the VT acoustics applies to the SGT acoustics as well.  We complement
this treatment by considering how and to what extent subglottal
acoustics plays a role in the vowel glide simulations reported above

\begin{figure} [!t]
  \centering
\hspace{1mm}
\includegraphics[scale=0.29]{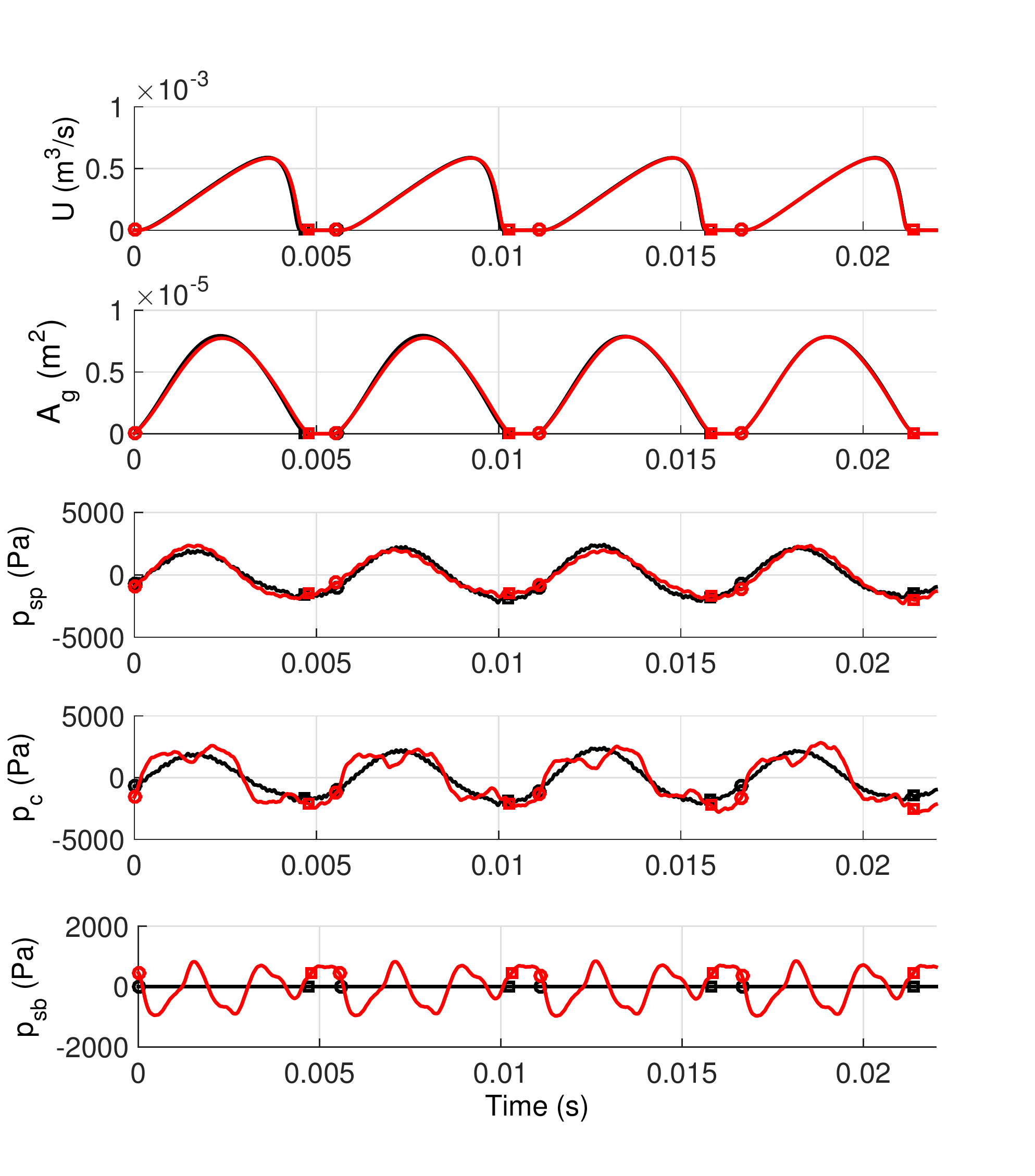} 
\includegraphics[scale=0.29]{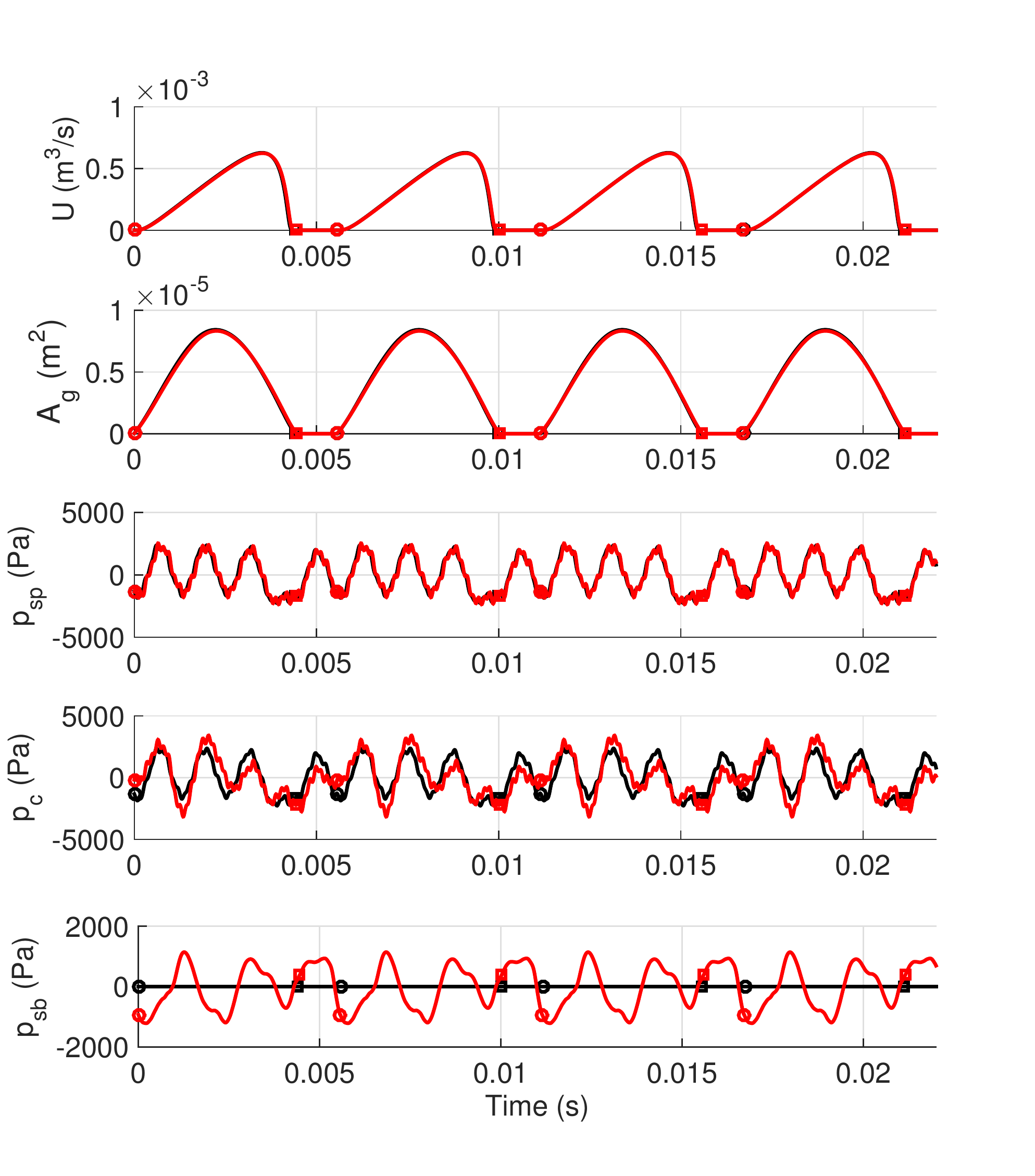} \\	
\caption{Left: Volume flow ($U$), glottal area ($A_g = \Delta W_1 h$), supraglottal
  pressure ($p_{sp}$) just superior to the vocal folds, counter
  pressure ($p_c$), and subglottal pressure just inferior to the vocal
  folds ($p_{sb}$) without SGT (black line) and with SGT (red line)
  for the vowel [\textipa{i}] at $f_o=180$ Hz.  Glottal closure are
  indicated by squares and openings by circles.  Right: Similar
  signals for the vowel [\textipa{\textscripta}].}
\label{SGT_ia}
\end{figure}

\RevisionText{Even though the acoustic termination at lungs is
  strongly resistive (see \cite{Lulich:SGTtissue:2015})}, significant
ringing \RevisionText{(i.e., oscillatory response to an abrupt change
  of a forcing)} takes place in the subglottal space. \RevisionText{In
  simulations, it is seen in the bottom panels of
  Figure~\ref{SGT_ia}}. Moreover, it has been verified by \emph{in
  vivo} measurements
\cite{Cranen:PM:1985,Cranen:SGF:1987,Neumann:2003,Zhang:SGAcoust:2005},
using physical models \cite{Austin:SGR:1997,Koike:1973}, and by
mathematically modelling the subglottal acoustics \cite{Ho:SGT:2011}
based on anatomic data of trachea and the progressively subdividing
system of bronchi and the alveoles \cite{Weibel:1963,Yeh:MHLA:1980}.
\RevisionText{A refined model for subglottal acoustic impedance was
  developed in \cite{Lulich:SGTtissue:2015} for the branching airway
  network in terms of transmission line theory, taking into
  consideration the contribution from yielding walls due to material
  parameters of cartilages and soft tissues.}
  
During the open phase, the inertia of the air column from bronchi up
to mouth opening is taken into account by $C_{iner}$ in
equation~\eqref{eq:vout}. At closure, the flow velocity $v_o$ drops to
zero, and a rarefaction pulse is formed above the vocal folds due to
air column inertia in the VT, and this is part of the acoustics
modelled by equation~\eqref{eq:webster}. Similarly, a compression
pulse is formed below the vocal folds, known as the ``water hammer''
in \cite{Sciamarella:WaterHammer:2009}. The subglottal resonator
equation~\eqref{eq:boundaries_sg} is mainly excited by the water
hammer. Both of these pulses can be seen in the supra- and subglottal
pressure signals $p_{sp}$ and $p_{sb}$ in Figure~\ref{SGT_ia}.

The water hammer is the most important component of subglottal
ringing, accompanied by its first echo that arrives back to vocal
folds after approximately $2 \, \mathrm{ms}$ delay. The delay corresponds to
the lowest subglottal formant between $500 \, \mathrm{Hz}$ and $600 \,
\mathrm{Hz}$ as reported in \cite{Cranen:SGF:1987} and \cite{Ho:SGT:2011}. The
first echo returns during the glottal closure at least if $f_o < 150
\, \mathrm{Hz}$ and the open quotient (OQ) of the pulse does not
exceed $50$ \%; see Figure~4 in \cite{Cranen:PM:1985} for measurements 
and Figure~12 in \cite{Ho:SGT:2011} for simulations. The echoes of the
water hammer pulse can be clearly seen in Figure~\ref{SGT_ia} as well
but now the first echo returns after the glottis has opened again due
to higher values of $f_o$ and OQ in these simulations.
 
The observations from simulations indicate that the subglottal
acoustics has an observable effect on glottal pulse waveform.  The
subglottal effect will get more pronounced when $f_o \to 
f'_{R1} = 500 \, \mathrm{Hz}$ which is the predefined frequency of the
first subglottal resonance.  This can be understood in terms of the
supraglottal behaviour shown in Figure~\ref{SGT_ia} since both the VT
and the SGT resonators have been realised similarly within the full
model. The sensitivity of the $f_o$ trajectory in the range
$\GlideFrequencyBand$ for the subglottal effect depends on the
magnitude of the SGT component of the counter pressure, regulated by
the parameter $c_3$ in equation~\eqref{eq:pc}. Considering the model
behaviour at supraglottal \ThirdRevision{resonance fractions} of
$f_{R1}$[\textipa{\textscripta}],
it is to be expected that the first subglottal \ThirdRevision{resonance 
fraction} $f'_{R1}/2$
should show up similarly.  This, indeed, happens if the coupling
constant $c_3$ in equation~\eqref{eq:pc} is large.

\subsection{Flow model}

The glottal flow described by equations~\eqref{eq:vout}--\eqref{eq:rtot}
contains terms representing \SecondRevision{the effect of viscosity 
in the glottis as well as pressure loss and recovery at the entrance and 
exit of the glottis, respectively.} Viscous pressure loss can easily be seen to be
significant by considering the glottal dimensions and viscosity of air
in the first term of equation~\eqref{eq:rtot}. It is clear from this
equation that the viscous losses dominate at least if the glottal
opening is small. 

The importance of \SecondRevision{entrance and exit effects} 
during parts of the glottal open phase can be seen, for example, by
comparing simulated volume velocities and glottal opening areas with
the experimental curves in Figure~3 in \cite{Berg:1957}, obtained from
a physical model of the glottis. In model simulations, leaving out
\SecondRevision{this transglottal pressure loss term} changes the
glottal pulse waveform significantly if other model parameters are
kept the same, as shown in Figure~3.7 in
\cite{Murtola:MVP:2014}. About half of the total pressure loss in
simulations is due to \SecondRevision{entrance and exit effects} at
the peak of opening of the glottis; see Figure~3.6 in
\cite{Murtola:MVP:2014}.  However, the behaviour of the simulated
$f_o$ trajectories over $f_{R1}$[\textipa{i}] does not change if the
\SecondRevision{transglottal pressure loss term} is removed. Then,
however, the vowel glide must be produced by different model parameter
values.

\section{\label{DiscSec} Discussion}

We have reported observations on the locking of simulated
$f_o$ glides on a resonance of the VT. The locking behaviour shows a
consistent time-dependent behaviour that is similar for rising and
falling glides. \SecondRevision{The $f_o$ jump at the beginning of the 
locking in rising glides and end of the locking in falling glides 
occurs together with and increased breathiness of phonation as 
characterised by open quotient $OQ$ and closing quotient $ClQ$. During 
the locking plateau, these parameters indicated an approximately steady 
change of phonation type.}

The locking takes place only at frequencies determined
by sub- or supraglottal resonances. \SecondRevision{Use of $p_{sub}$ as a 
secondary control parameter for the glides ensure that the main cause for 
changes in $OQ$ and $ClQ$ is the acoustic loading.} By modifying the strength of the
acoustic feedback (i.e., the parameter $Q_{pc}$ in equation~\eqref{eq:pc})
and vocal fold tissue losses (i.e., the parameter $\beta$), the locking
tendency at $f_{R1}$[\textipa{i}] may be modulated from non-existent
(where both $Q_{pc}$ and $\beta$ have low values) to extreme locking
at $f_{R1}$[\textipa{i}] without release (where $Q_{pc}$ and/or $\beta$ have
exaggeratedly large values); see Figure~\ref{qual_b_Q}.  By decoupling
secondary components from the simulation model, 
the locking behaviour at $f_{R1}$[\textipa{i}] remains the same, even
though the model parameter values required for the desired glottal
waveform change. We conclude that the simulation results on vowel
glides reported above
reflect the model behaviour in a consistent and robust manner.

Vowel glides observed in test subjects are another matter. Any model
is a simplification of reality, and there is a catch in assessing the
role of unmodelled physics: a proper treatment would require the
modelling of it. Short of this, we discuss these aspects based on
literature, model experiments, and reasoning by analogy.

\subsection{Acoustics}

Viscosity of air has not been taken into account in the acoustics
model though a measurable effect is likely take place in narrow parts
of the VT or SGT. Resulting attenuation can be treated by adding a
dissipation term of Kelvin--Voigt type to equations~\eqref{eq:webster}
and \eqref{eq:boundaries_sg}. For a constant diameter waveguide, the
term is proportional to $\mu \psi_{sst} / c^2$. Adding viscosity
losses will widen and lower the resonance peaks of Webster's
resonators (i.e., lower their Q-value), with a slight change in the
centre frequencies. An analogous effect can be studied by increasing
the tissue dissipation parameter $\alpha_1$ in
equation~\eqref{eq:webster} \RevisionText{(or $\alpha_2$ in
equation~\eqref{eq:boundaries_sg})} to a very high value which has been
observed not to change the conclusions on vowel glide simulations.
\ThirdRevision{Similarly, the overall acoustic
  resistance of the VT has no qualitative effect which can be seen in
  \cite{Aalto:Interaction:2011} where the modal locking was observed
  for vowel [\textipa{\oe}] at the lowest resonance $647 \,
  \mathrm{Hz}$ despite the fact that the anatomic geometry of
         [\textipa{\oe}] has a much wider flow channel than that of
         [\textipa{i}].}

The SGT modelling by the horn is a crude simplification of the
fractal-like lower airways and lungs.  The network structure of the
subglottal model in \cite{Ho:SGT:2011} could be replicated by
interconnecting a large number of Webster's resonators, each modelled
by equation~\eqref{eq:boundaries_sg}. The resulting transmission graph
is a passive dynamical system by Section~5 in \cite{Aalto:PBCS:2013},
but it is not clear how to write an efficient FEM solver for such
configurations. 

The model proposed in \cite{Ho:SGT:2011} as well as the
transmission graph approach are likely to produce the correct
resonance distribution and frequency-dependent energy dissipation rate
at the lung end without tuning.  The horn model does require tuning of
the horn opening area and the boundary condition on it in order to get
realistic behaviour on the lowest subglottal resonance $f'_{R1} = 500 
\, \mathrm{Hz}$. Doing so freezes all the higher subglottal resonances 
at fixed positions, e.g., $f'_{R2} = 1 \,\mathrm{kHz}$. The 
branching subglottal models given in Figure~8 in \cite{Ho:SGT:2011} 
have the second subglottal resonance between $1.3 \, \mathrm{kHz}$ and 
$1.5 \, \mathrm{kHz}$. \RevisionText{It was observed in
  \cite{Lulich:SGTtissue:2015} that the soft tissues introduce an
  additional resonance to the subglottal system that is lower than the
  first subglottal formant $F'_{1}$ due to air column dynamics. The
  is no obvious way how a horn model could be used to accommodate such
  a resonance at $\approx 350 \, \mathrm{Hz}$ due to the yielding wall
  dynamics.}

Based on the observations on the simulated vowel glides,
it seems convincing that the overall subglottal effect on the
fundamental frequency $f_o$ is insignificant for vowel glides within
$\GlideFrequencyBand$ that is over $100\,\textrm{Hz}$ away from
$f'_{R1}$. However, the subglottal effect is certainly discernible in
waveforms as in Figure~\ref{SGT_ia}, but the effect of higher
subglottal \ThirdRevision{resonances} $f'_{R2}, f'_{R3}, \ldots$ cannot 
be seen even there. In current simulations of female phonation, the vocal fold
mass-spring system has its mechanical resonances at approximately $
150 \, \mathrm{Hz}$, which acts as a low-pass filter for subglottal
excitation in higher frequencies. The same conclusions are likely to
hold when using a more complicated subglottal resonator geometry with
one caveat: a graph-like subglottal geometry has lots of cross-mode
resonances that affect the subglottal acoustic impedance in other ways
than just moving the pole positions.

\RevisionText{Also the DC-component of the glottal flow loads the
  acoustic resonators in equations~\eqref{eq:boundaries} and
  \eqref{eq:boundaries_sg}. If we use $v_{ac}(t) = v_0(t) -
  \frac{1}{T}\int_{t - T}^t{v_0(\tau) \, d\tau}$ with $T = 2/f_o$
  instead of $v_0$ as input to the resonator equations, only
  negligible effects are observed in simulated stable waveforms. There
  are more pronounced effects in the beginnings of simulated phonation
  when a stable waveform has not yet developed.}

\subsection{Vocal fold geometry and glottal flow}

The idealised vocal folds geometry shown in Figure~\ref{fig:model}
(right panel) leads to a particularly simple expression for the
aerodynamic force in equation~\eqref{eq:aeroforces}.  Replacing the
sharp peaks by flat tops in Figure~\ref{fig:model} (but keeping the
same glottal gap $g$ at rest) results in phonation that has typically
lower open quotient (OQ) whereas the original wedge-like geometry
produces more often phonation where the glottis does not
close. \ThirdRevision{This change makes it easier to adjust the
  parametrisation of the model to obtain some phonation targets. In
  particular, the glottal loss parameter $k_g$ can be set closer to
  more commonly used, somewhat larger values since the model geometry
  becomes more similar to the geometries used in related literature.}

\ThirdRevision{Another aspect involving the aerodynamic force on vocal
  fold structures is associated with the hydrostatic pressure
  reference level in vibrating tissues. This level is denoted by
  $p_{ref}$, and it is expected to satisfy $p_{ref} \leq p_{sub}$. If
  the air pressure between the vocal folds were equal to $p_{ref}$,
  then the vocal fold wedges would not be accelerated by the pressure
  difference.  In equation~\eqref{eq:int2}, we use $p_{ref} =
  p_{sub}$, and we always have $p(x,t) - p_{sub} \leq 0$. For this
  reason, the effect of the aerodynamic force is always trying to
  close the glottis in this case. For small flow velocities $V(x,t)$,
  using $p(x,t) - p_{ref}$ with $p_{ref} < p_{sub}$ in
  equation~\eqref{eq:int2} would give the following outcome: the
  driving pressure $p_{sub}$ would push the vocal folds open more
  strongly than the aerodynamic force would pull them
  close. Unfortunately, there is no obvious way to determine the true
  magnitude of $p_{ref}$ as it is an outcome of dynamic pressure
  equalisation processes related to $p_{sub}$ and the additional
  partial pressure due to haemodynamics in tissues. Using a tuned
  value of $p_{ref}$ instead of $p_{sub}$ in equation~\eqref{eq:int2}
  would be desirable, e.g., in phonation onset simulations; in
  particular, if $g = 0$ where using $p_{ref} = p_{sub}$ would not
  start a phonation cycle at all.}




The glottal flow has been studied extensively since 1950's.  Compared
to the flow model given above,
physiologically more faithful glottal flow solvers have been proposed
in, e.g., \cite{Pelorson:FlowSepar:1994},
\cite{Titze:MaxPowerTrans:2002} and \cite{Erath:ThPField:2011}; see
also \cite{Birkholz:Turbulence:2007}, \cite{Ishizaka:1972},
\cite{Punvcochavrova:2010}, and \cite{Berg:1957}.
As pointed out in \cite{Punvcochavrova:2010}, more sophisticated flow
models are challenging to couple to acoustic resonators since the
interface between the flow-mechanical (in particular, the turbulent)
and the acoustic components is no longer clearly defined. 

Flow separation and Coand\u{a} effect during the diverging phase of
the phonatory cycle (which obviously cannot occur in wedge-like
geometry of Figure~\ref{fig:model}) have been studied in
\cite{Erath:ThPField:2011}, \cite{Erath:AsymFlow:2010} and
\cite{Pelorson:FlowSepar:1994} using boundary layer theory and
physical model experiments. The boundary layer leaves the vocal fold
surfaces at the time-dependent flow separation point, say $x_s$,
forming a jet which extends downstream into supraglottal space.  Thus,
the vocal folds ``stall'' at $x_s$, and the aerodynamic force on them
is greatly diminished; see Section~IV in \cite{Erath:ThPField:2011}
where the vocal fold model is from
\cite{Steinecke:Bifurcations:1995}. Similarly, the viscous pressure
loss equation~(A7) in \cite{Pelorson:FlowSepar:1994} depends only on
the upstream part of glottis that ends at $x_s$. Simplifying
assumptions on the vocal fold geometry \cite{Pelorson:FlowSepar:1994}
are required for computing $x_s$, and the result is sensitive to the
geometry which makes it challenging to model.

Turbulence in supraglottal space is a spatially distributed acoustic
source, and it does not provide a scalar flow velocity signal for
boundary control as $v_0$ in equation~\eqref{eq:boundaries}.  The
supraglottal jet may even exert an additional aerodynamic force to
vocal folds that would not be part of the acoustic counter pressure
$p_c$ from the acoustic resonators.  Turbulence in VT constrictions is
the primary acoustic source for unvoiced fricatives, and many such
sources have been modelled separately in, e.g., 
 \cite{Birkholz:Turbulence:2007}.
Much of the turbulence noise energy lies above $4\, \mathrm{kHz}$ but
Webster's model equation~\eqref{eq:webster} is an accurate description of
VT acoustics only below $4\, \mathrm{kHz}$ due to the lack of
cross-modes \cite{Vampola:VTResFEM:2013,Vampola:FEMVT:2011}.
This fact speaks against the wisdom of including turbulence noise in the
proposed model.

The proposed phonation model 
treats flow-mechanic\-al and acoustic components using separate
equations, and we conclude that this paradigm is not conducive for
including the advanced flow-mechanical features discussed
above. Instead, phonation models based on Navier--Stokes equations
would be a more appropriate framework.

\section{Conclusions}

We have presented a model for vowel production, based on (partial)
differential equations, that consists of submodels for glottal flow,
vocal folds oscillations, and acoustic responses of the VT and
subglottal cavities. The model has been originally designed as a
tunable glottal pulse source for a high-resolution VT acoustics
simulator that is based on the 3D wave equation and VT geometries
obtained by MRI as explained in \cite{Aalto:MesTec:2014,Kivela:MT:2015}.
The model has found applications as a controlled source of synthetic
vowels that are needed in, e.g., developing speech processing
algorithms such as the inverse filtering
\cite{Alku:IAIF:1992,Alku:IFReview:2011}.

In this article, the model was used for simulations of rising and
falling vowel glides of [\textipa{\textscripta, i}] in frequencies
that span one octave $\GlideFrequencyBand$. This interval contains the
lowest VT resonance $f_{R1}$ of [\textipa{i}] but not that of
[\textipa{\textscripta}]. Perturbation events in simulated vowel
glides were observed at VT acoustic resonances, or at some of their
fractions but nowhere else. The fundamental
frequency $f_o$ of the simulated vowel was observed to lock to
$f_{R1}$[\textipa{i}] but similar locking was not seen at any of the
\ThirdRevision{resonance fractions}. \SecondRevision{The locking events were accompanied by 
changes in the phonation: increased breathiness below and partially at 
the locking frequency and steady change in breathiness during most of the 
lock.} Such modal locking event takes place only when the
acoustic feedback from VT to vocal folds is present, and then it has a
characteristic time-dependent behaviour. A large number of simulation
experiments were carried out with different parameter settings of the
model to verify the robustness and consistency of all observations.

To what extent do the simulation results validate the proposed model?
The model produces perturbations of the glottal pulse both at VT
resonances and at some of the VT \ThirdRevision{resonance fractions}. Of the former, a wide
existing literature was reviewed in Introduction.
Observations on the subformant perturbations in speech have not been
reported, to our knowledge, in experimental literature. There is a
particular temporal pattern of locking in simulated perturbations at
$f_{R1}$[\textipa{i}] as explained in Figure~\ref{qual_b_Q} (left
panel). \ThirdRevision{Although pulse based $f_o$ trajectories are
  rarely shown in literature, a similar} pattern can be seen in the
speech spectrograms given in Figure~5 in
\cite{Titze:NonlinCouplingExp:2008}, \ThirdRevision{and} Figure~4 in
\cite{Tokuda:RegisterTransitions:2010}, as well as in vowel glide
samples in the data set of the companion article
\cite{A-M-V:MLBVFVTOPII}. \SecondRevision{A similar locking behaviour
  can also} \ThirdRevision{be seen in simulated spectrograms in
  Figure~6 in \cite{Ruty:WaveguideLength:2008} and Figures~13 and 14
  in \cite{Titze:NonlinCouplingTh:2008}, and it can also}
\SecondRevision{be interpreted to lie behind the experimental results
  shown in Figures~10b~and~13b of \cite{Ruty:2007}.}  The glottal flow
and area simulations in Figure~\ref{SGT_ia} are remarkably similar
with the experimental data presented in Figures~4-7 in
\cite{Cranen:PM:1985}, the signals produced by different numerical
models (see Figures~14a-14c in \cite{Ishizaka:1972}, Figures~8 and 10
in \cite{Zanartu:AL1MM:2007}, Figures~10--11 in \cite{Ho:SGT:2011},
Figure~6 in \cite{Titze:MaxPowerTrans:2002}, Figure~5 in
\cite{Titze:Parametrization:1984}), and the glottal pulse waveforms
obtained by inverse filtering in, e.g., Figures~10--13 in
\cite{Alku:IAIF:1992}, Figures~5.3, 5.4, and 5.17 in
\cite{Pulakka:MT:2005}, \cite{Aalto:LF:2009}, and Figures~3 and 6 in
\cite{Alku:IFReview:2011}.

\SecondRevision{The simulation model does not include} the neural 
control actions on the vocal fold structures \SecondRevision{or dynamic} 
modifications of the VT geometry.  There is
also a significant control action affecting the subglottal pressure
and it has been used as a control variable in
equations~\eqref{RisingScalingEq}--\eqref{FallingScalingEq} for glide
productions.  In humans, neural control actions are part of feedback
loops, of which some are auditive, and some others operate directly
through tissue innervation and the central nervous system. So little
is known about these feedback mechanisms that their explicit mathematical 
modelling seems \SecondRevision{infeasible}. Instead, the model parameters
for simulations are tuned so that the simulated glottal pulse waveform
corresponds to experimental speech data. \SecondRevision{Despite these
simplifications the model appears to be sufficiently detailed to replicate
the observations found in literature.}


\section*{Acknowledgements}
  
The authors were supported by the Finnish graduate school in
engineering mechanics, Finnish Academy project Lastu 135005, 128204,
and 125940; European Union grant Simple4All (grant no. 287678), Aalto
Starting Grant 915587, and {\AA}bo Akademi Institute of
Mathematics. The authors would like to thank the four anonymous
reviewers in 2013 and 2016 for comments leading to many improvements
of the model.




\end{document}